\documentclass[graphics,floatfix, footinbib,tightenlines,nobibnotes, aps, prb, twocolumn]{revtex4-1}
\usepackage{amsmath,amssymb}
\usepackage{graphicx}
\usepackage{dcolumn}
\usepackage{bm}
\usepackage{braket}
\usepackage{subcaption}
\usepackage{verbatim}
\usepackage{float}
\usepackage{bbold}
\usepackage{color}
\usepackage{xcolor}
\usepackage{relsize}
\usepackage{amsthm}
\usepackage{enumerate}
\usepackage{soul,xcolor}
\usepackage[T1]{fontenc}
\usepackage[colorlinks=true ,urlcolor=blue,urlbordercolor={0 1 1}]{hyperref}

\newcommand{\beq}{\begin{eqnarray}}
\newcommand{\eeq}{\end{eqnarray}}

\newcommand{\bsp}{\begin{split}}
\newcommand{\esp}{\end{split}}

\newcommand{\be}{\begin{equation}}
\newcommand{\ee}{\end{equation}}
\newcommand{\mf}{\mathbf}

\begin{document}

\setstcolor{red}

\title{Bridging Hubbard Model Physics and Quantum Hall Physics in Trilayer Graphene/h-BN moir\'e superlattice}
\author{
Ya-Hui Zhang}
\affiliation{Department of Physics, Massachusetts Institute of Technology, Cambridge, MA, USA
}
\author{T. Senthil}
\affiliation{Department of Physics, Massachusetts Institute of Technology, Cambridge, MA, USA
}

\date{\today}

\begin{abstract}
The moir\'e superlattice formed by  ABC stacked trilayer graphene aligned with a hexagonal boron nitride substrate (TG/h-BN) provides an interesting system where both the bandwidth and the topology can be tuned by an applied  perpendicular electric field  $D$ . Thus the TG/h-BN system can  simulate both Hubbard model physics   and nearly flat Chern band physics   within one sample.  We derive lattice models for both signs of $D$ (which controls the band topology) separately through explicit Wannier orbital construction and mapping of Coulomb interaction.  When the bands are topologically trivial,  we discuss  possible candidates for Mott insulators at integer number of holes per site (labeled as $\nu_T$). These include both broken symmetry states and quantum spin liquid insulators which may be particularly favorable in the vicinity of  the Mott transition. We propose feasible experiments to study carefully the bandwidth tuned and the doping tuned Mott metal-insulator transition at both $\nu_T=1$ and $\nu_T=2$.     We discuss the interesting  possibility of probing experimentally a bandwidth (or doping) controlled  continuous Mott transition  between a Fermi liquid metal and a  quantum spin liquid  insulator. Finally  we also show that the system has a large valley Zeeman coupling  to a small out-of-plane magnetic field, which can be used to control the valley degree of freedom.  
\end{abstract}

\pacs{Valid PACS appear here}
\maketitle

\section{Introduction}

Recently moir\'e superlattices in twisted Van der Waals heterostructures have been shown to realize several strongly correlated systems with high tunability\cite{spanton2018observation,cao2018correlated,cao2018unconventional,chen2018gate,yankowitz2018tuning}. Correlated insulators and superconductors have been reported experimentally in twisted bilayer graphene\cite{cao2018correlated,cao2018unconventional,yankowitz2018tuning} and in   ABC stacked graphene/hexagonal boron nitride (TG/h-BN)\cite{chen2018gate}.  In this paper we focus on the  TG/h-BN system. 

\begin{figure}
\centering
\includegraphics[width=0.45\textwidth]{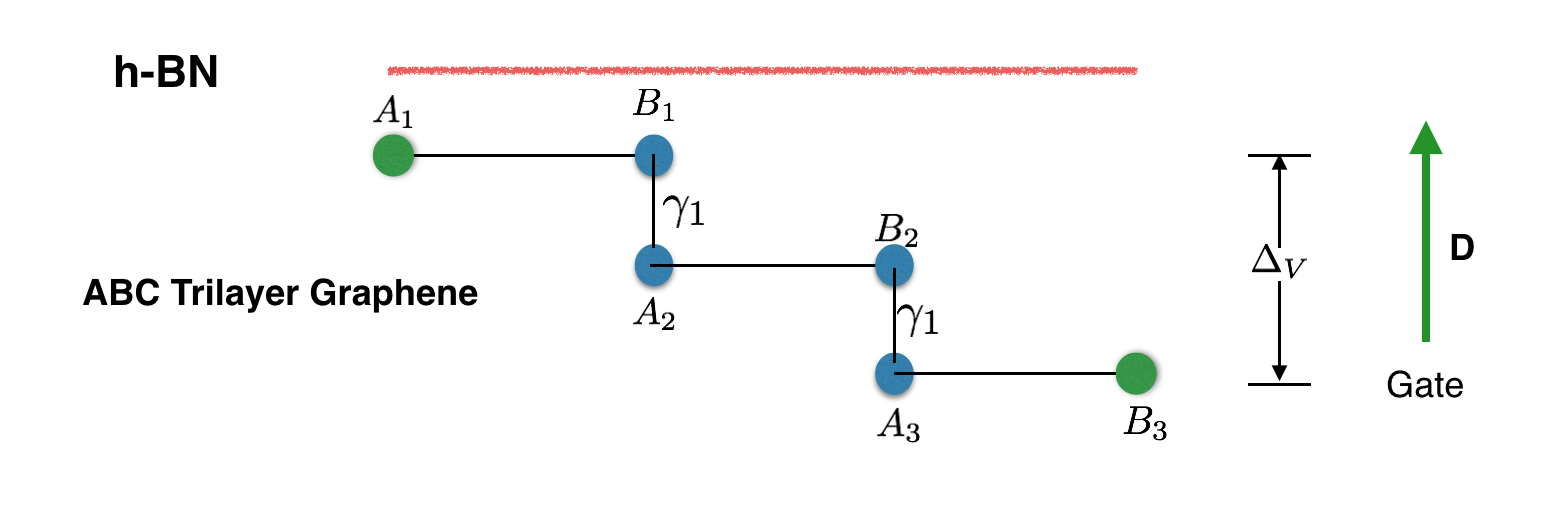}
\caption{Illustration of the ABC stacked trilayer graphene/h-BN system. We assume the  h-BN layer on top is nearly aligned with the graphene layers while the one at the bottom is not aligned.   $A$ and $B$ refer to the 2 sublattices in each of the graphene layers. Due to the large dimerization term $\gamma_1\approx 400$ meV, only $A_1$ and $B_3$ should be kept at low energy, forming a two-component spinor. A vertical electric field gives an energy difference $\Delta_V$ for electrons between the top and the bottom graphene layer. The aligned h-BN layer provides a moir\'e superlattice potential which folds the original large Brillouin Zone to a small moire Brilloiun zone (MBZ).}
\label{fig:system_illustration}
\end{figure}

Bandwith\cite{chen2018gate} and even band topology\cite{zhang2018moir} can be tuned by an applied perpendicular electric field $D$  in   TG/h-BN. The displacement field $D$ provides an energy difference  $\Delta_V$ for electrons between the top and the bottom graphene layer, as illustrated in Fig.~\ref{fig:system_illustration}. For $\Delta_V<0$ (this convention assumes that the h-BN layer on top is nearly aligned with the TG), the bands of the two valleys have zero Chern number while for $\Delta_V>0$ they have non-zero Chern numbers $C=\pm 3$\cite{zhang2018moir,chittari2018gate}. Correlated insulators are found at $\nu_T=1$ and $\nu_T=2$ for the valence band of TG/h-BN at large $|\Delta_V|$ \cite{chen2018gate}, where $\nu_T$ is defined as the total density of holes per moir\'e unit cell. When $\Delta_V>0$,  physics similar to quantum Hall systems may be realizable.  For trivial narrow bands that obtain when  $\Delta_V<0$,   the physics is expected\cite{po2018origin} to be governed by an anisotropic $SU(4)$ Hubbard model (with small anisotropies) at leading order. Therefore TG/h-BN  offers an experimental system where both Hubbard model physics and quantum Hall like physics can be simulated by simply switching the gate.

In this paper we describe several new aspects of the physics of TG/h-BN with a focus on the topologically trivial side ($\Delta V < 0$).  We obtain an explicit  interacting lattice   model and estimate its parameters using the
continuum description of the moire band structure\cite{bistritzer2011moire}.  We use this lattice model to discuss the physics both deep in the correlated insulator regime and in the regime close to the Mott metal-insulator transition. We highlight the opportunities presented by this system to tunably study both the bandwidth tuned and doping tuned Mott transitions. We propose a number of transport experiments that can  probe the Mott transition.  We also present some new results on the topological bands that obtain for $\Delta_V > 0$. 

 For $\Delta_V<0$, we build Wannier orbitals following the standard approach, and explicitly  construct an effective tight-binding model. We  project the Coulomb interactions to determine the effective  interactions in the lattice model.  The result  is a spin-valley extended Hubbard model with Hund's couplings as much smaller perturbations. The $SU(4)$ symmetry from the spin-valley degrees of freedom is mainly broken by a valley-contrasting flux in the hopping.    Based on this model, we argue  that the insulators found in the experiment should be understood as standard Mott insulators with charge frozen by Hubbard $U$, in contrast to the nesting scenario in Ref.~\onlinecite{zhu2018antiferro}.  In the limit of a nearly flat band, we argue that the insulator should be a {\em ferromagnet} for both $\nu_T=1$ and $\nu_T=2$. For intermediate strength interactions, quantum spin liquids phases are  promising candidates. In the vicinity of the Mott transition, a natural candidate is a spin liquid with neutral fermi surface coupled to an emergent $U(1)$ gauge field. 
  
The Mott metal-insulator transition\cite{imada1998metal} is a fundamental phenomenon in condensed matter physics. Graphene moire systems like TG/h-BN offer a wonderful opportunity to controllably tune through the transition and explore its properties. It has long been appreciated that there are a number of distinct routes to the Mott transition in correlated solids. We describe distinctive signatures - visible in feasible  experiments on TG/h-BN -   of  some of these distinct routes.  Most striking is the possibility\cite{senthil2008theory} of a bandwidth tuned continuous quantum critical Mott transition from the Fermi liquid metal to a spin liquid with a neutral Fermi surface. We show how to  explore such a continuous Mott transition through simple transport experiments:  a universal jump of residual resistivity at the critical point and  Shubnikov-deHaas oscillations even inside the Mott insulator.  Besides, we also discuss the possibility of a doping controlled continuous metal-insulator transition (DMIT) between the above two phases. Interestingly we find that the existing experimental data in Ref.~\onlinecite{chen2018gate} may already have signatures of such a doping tuned continuous metal-insulator transition close to the filling $\nu_T=2$.  
 
 Finally, we show that there is a large valley Zeeman coupling with averaged g factor $g \sim 54$. Therefore, a small out of plane magnetic field can polarize the valley and lead to a spin $1/2$ model. We discuss some consequences of this phenomenon. 

 For the  topologically non-trivial $\Delta_V  > 0$ side, valley preserving localized Wannier orbitals are impossible because of the non-zero Chern number $C=\pm 3$. Related but distinct  Wannier obstructions have also been discussed in the context of the twisted bilayer graphene system\cite{po2018origin,zou2018clarify,ahn2018failure,song2018all}.  The Wannier obstruction for the $\Delta_V>0$ TG/h-BN system can not be removed by adding trivial bands and is therefore different from the fragile topology of the twisted bilayer graphene system\cite{po2018faithful}.   Following a similar treatment of twisted bilayer graphene in Ref. \onlinecite{po2018origin} we build a two orbital model on the triangular lattice, though the valley charge operator  is not  a sum of  on-site terms.  As argued in our previous work\cite{zhang2018moir} the $\Delta_V>0$ side is promising to realize a quantum Anomalous Hall insulating state with strong interactions at  $\nu_T=1$. At fractional fillings, fractional quantum Anomalous Hall states may also be possible. The model derived in the present paper may in the future aid quantitative theoretical and numerical studies of these phenomena.

\section{Lattice Model For $\Delta_V<0$ Side: spin-valley Hubbard Model \label{section:lattice_model_trivial}}

Band structures of TG/h-BN  were  calculated in Ref.~\onlinecite{zhang2018moir} using a continuuum model. 
An important  feature, as demonstrated experimentally in Ref. \onlinecite{chen2018gate},  is that the band width can  simply be tuned by the perpendicular displacement field $D$ (equivalently the potential difference $\Delta_V$).   More details of the band structure can be found in Appendix.~\ref{appendix:band_structures}.  Here we will use the results on the band structure to build an interacting  lattice model  with a focus on the topologically trivial $\Delta_V<0$  side. 

When  $\Delta_V<0$, the valence band of each valley has zero Chern number and exponentially localized Wannier orbital on triangular lattice can be constructed for each valley separately. Following the methods in Appendix.~\ref{appendix:lattice_models}, we  derived an interacting  triangular lattice model which we describe below.  At each site $\mathbf x$ of the lattice there are 4 single particle states corresponding to 2 spin and 2 valley degrees of freedom. We work in the hole picture. We 
write the corresponding hole  destruction operator as $\psi_{a,\sigma}(\mathbf x)$ where $a = \pm$ is the valley index and $\sigma = \uparrow, \downarrow$ is the spin index.  

Microscopically the system has symmetries of charge conservation, spin rotation, and time reversal. The latter acts by flipping the two valleys\footnote{It is convenient to define time reversal  without flipping the spin.  We are free to combine this with a spin rotation to obtain a modified time reversal operation which flips both spin and valley.} . : 
\be
\label{treversal}
{\cal T}: \psi_{a,\sigma}(\mathbf x) \rightarrow \left(\tau^x \right)_{ab} \psi_{b,\sigma}(\mathbf x)
\ee

For a large period moire structure (super)-lattice translations are an excellent symmetry as is a $C_3$ rotation (about a triangular site) which acts as 
\be
\label{c3}
{\cal C}_3 : \psi_{a,\sigma}(\mathbf x) \rightarrow \psi_{a,\sigma}(\mathbf x')
\ee
where $\mathbf x'$ is the site to which $\mathbf x$ is taken by the $C_3$ rotation.   Further, to an excellent approximation, the number of electrons within each valley is independently conserved. There is a corresponding valley charge $U(1)$ symmetry. Finally within the continuum model there is a mirror reflection symmetry which also interchanges the two valleys (see Appendix.~\ref{appendix:band_structures}): 
\be
\label{M}
{\cal M}: \psi_{a,\sigma}(\mathbf x) \rightarrow \left(\tau^x \right)_{ab}\psi_{b,\sigma}(\mathbf x')
\ee
where $\mathbf x'$ is generated from $\mathbf x$ by a mirror reflection plane passing through $\mathbf{a_1} +\mathbf{a_2}$ where $\mathbf{a_1}=a_M(1,0)$ and $\mathbf{a_2}=a_M(\frac{1}{2},\frac{\sqrt{3}}{2})$ are two unit vectors for the triangular lattice. 

Note that there is no microscopic $C_6$,  and hence $C_2$ symmetry. If present, $C_2 T$ will forbid any non-zero Berry curvature at generic points in the MBZ. However, there exist non-zero Berry curvature close to the $\Gamma$ point and the MBZ boundary\cite{zhang2018moir} though their sum cancels for the $\Delta_V<0$ side. In the next sub-section we will also show that there is a large out of plane orbital magnetic moment $\mathbf{m}(\mathbf k)$ at each momentum $\mathbf k$, which can not be compatible with the existence of both time reversal and $C_6$ symmetry.

Below we will derive a lattice model for the active bands. In the non-interacting limit, despite its absence as a microscopic symmetry,  the lattice tight-binding model  is symmetric under a  $C_6$ rotation (about a triangular site) which acts as 
\be
\label{c6}
{\cal C}_6 : \psi_{a,\sigma}(\mathbf x) \rightarrow \left(\tau^x \right)_{ab}\psi_{b,\sigma}(\mathbf x')
\ee
where $\mathbf x'$ is the site to which $\mathbf x$ is taken by the $C_6$ rotation. Thus $C_6$ flips the two valleys. This symmetry will be broken by interaction terms. However we will see that the part of the interaction that breaks $C_6$ is much smaller than other terms. Hence $C_6$ will be a good approximate symmetry of the effective lattice model though it is not a microscopic symmetry.

Using these symmetries, the lattice tightbinding model can be written
 \begin{align}
 	H_K&=-\sum_{\mathbf x;\sigma}\sum_{m,n}t(m,n)\psi^\dagger_{+\sigma}(\mathbf x +m \mathbf{a_1}+n\mathbf{a_2})\psi_{+\sigma}(\mathbf x)+h.c.\notag\\
 	&-\sum_{\mathbf x;\sigma}\sum_{m,n}t^*(m,n)\psi^\dagger_{-\sigma}(\mathbf x +m \mathbf{a_1}+n\mathbf{a_2})\psi_{-\sigma}(\mathbf x)+h.c.
  \label{eq:kinetic_trivial}
 \end{align}
  $\pm$ is valley index and $\sigma=\uparrow,\downarrow$ is spin index.
We need only  the following hopping terms: $t_1=t(1,0)$, $t_2=t(1,1)$, $t_3=t(1,2)$ and $t_4=t(2,0)$, and other terms that can be generated by $C_6$ rotation and $M$ reflection symmetry.

We list tight binding  parameters for different $\Delta_V$ in Table.~\ref{table:tight_binding_parameters}. A key feature\cite{po2018origin} allowed by the symmetries is that within a single valley there is no time reversal, and hence there can be a non-zero flux through each triangular plaquette. However this flux must be opposite on neighboring plaquettes. From the explicit calculations of the tightbinding parameters  we see that the staggered flux in one triangle for each valley is about $0.5 \pi -2\pi$ in the regime $\Delta_V <-25$ meV. Such a valley contrasting flux strongly breaks the spin-valley $U(4)$ symmetry\footnote{This is a combination of total charge $U(1)$ transformation and the spin-valley $SU(4)$ rotation.}  down to $U(2)_+\times U(2)_-$. Here $U(2)_a$ means an independent $SU(2)$ spin rotation combined with $U(1)$ transformation  for each valley $a$.  As we   show in the next section, this $U(4)$ symmetry breaking term  will be inherited in the spin-valley model of the Mott insulator through   super-exchange.

\begin{figure}
\centering
\includegraphics[width=0.45\textwidth]{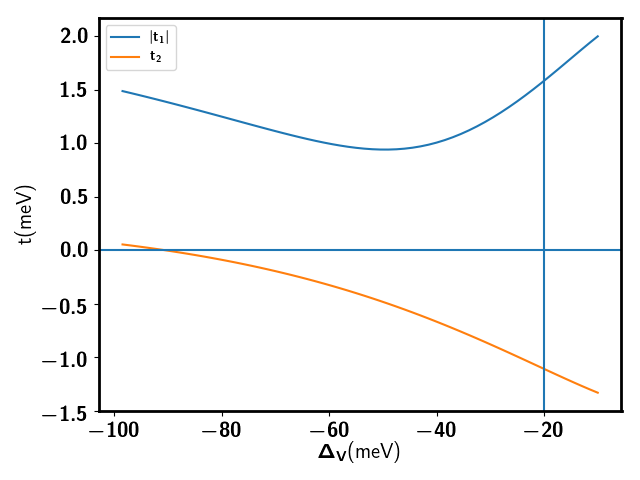}
\caption{Magnitude of the nearest neighbor hopping $|t_1|$ and the next-nearest neighbor hopping $t_2$. $t_2$ has no imaginary part because of the Mirror reflection symmetry. The phase of $t_1$ is shown in Fig.~\ref{fig:hopping_flux}. The vertical line labels $\Delta_V=-20$ meV where the bandwidth is equal to the Hubbard $U$: $W\approx U\approx 25$ meV.}
\label{fig:hopping_magnitude}
\end{figure}

\begin{figure}
\centering
\includegraphics[width=0.45\textwidth]{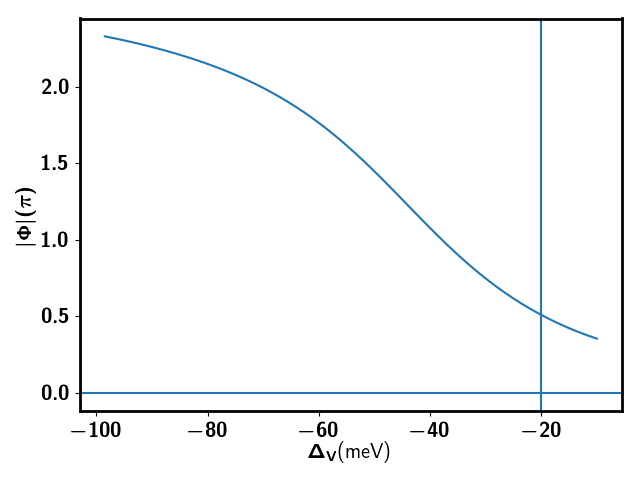}
\caption{The flux $|\Phi|$ of each triangle from the nearest neighbor hopping. For each triangle, two valleys experience opposite $\Phi$. For each valley, $\Phi$ changes sign under $C_6$ rotation. The vertical line labels $\Delta_V=-20$ meV where the bandwidth is equal to the Hubbard $U$: $W\approx U\approx 25$ meV. For the Mott insulating regime at $\Delta_V <-25$ meV, we expect a large valley contrasting flux $|\Phi|\sim 0.5\pi-2\pi$ trhough each triangle. Such a flux breaks $SU(4)$ symmetry, which is inherited in the spin model for the Mott insulator through the super-exchange term.}
\label{fig:hopping_flux}
\end{figure}

\begin{table}[h]
\centering
\begin{tabular}{c|c|c|c|c|}

\hline
$\Delta_V$&$t_1$&$t_2$&$t_3$&$t_4$\\
\hline
$-100$&$1.505e^{i 0.780\pi}$&$-0.063$& $0.046e^{-i 0.544\pi}$& $0.323e^{-i 0.292\pi}$\\
\hline
$-70$&$1.113e^{i 0.664 \pi}$&$-0.195$& $0.089 e^{-i 0.305\pi}$& $0.407 e^{-i 0.396\pi}$\\
\hline
$-50$&$0.941e^{i 0.482 \pi}$&$-0.482$& $0.158 e^{-i 0.181\pi}$& $0.478 e^{-i0.487\pi}$\\
\hline
$-30$&$1.227e^{i0.249 \pi}$&$-0.879$& $0.267 e^{-i 0.100\pi}$& $0.610 e^{-i0.599\pi}$\\
\hline
$-20$&$1.583e^{i0.169 \pi}$&$-1.108$& $0.323 e^{-i 0.069\pi}$& $0.732 e^{-i0.653\pi}$\\
\hline
$-10$&$1.998e^{i0.118 \pi}$&$-1.330$& $0.4363e^{-i 0.035\pi}$& $0.905 e^{-i0.692\pi}$\\
\hline
\end{tabular}
\caption{Tight binding parameters for $\Delta_V<0$ side. Both $\Delta_V $ and $t$ are in units of meV.}
\label{table:tight_binding_parameters}
\end{table}

To obtain the  interaction we start with the (screened) Coulomb interaction and project it on to the active valence bands, as explained in Appendix.~\ref{appendix:lattice_models}. We find 
\begin{align}	
	H_V&=\frac{U}{2} \sum_i n_i^2+g_1 U \sum_{\langle ij \rangle}n_i n_j \notag\\
	&-\frac{2 g_h U}{2}  \sum_{\langle ij \rangle}\sum_{a_1a_2;\sigma_1\sigma_2} \psi^\dagger_{i;a_1\sigma_1}\psi_{i;a_2\sigma_2} \psi^\dagger_{j;a_2\sigma_2}\psi_{j;a_1\sigma_1} \notag\\
	&+\frac{J'_H}{2} \sum_{\langle ij \rangle}\sum_{\sigma_1\sigma_2}\left(\psi^\dagger_{i;+\sigma_1}\psi_{i;-\sigma_2} \psi^\dagger_{j;-\sigma_2}\psi_{j;+\sigma_1} +h.c.\right)\notag\\
  &-J_{H} \sum_i \left(\frac{1}{4}n_{+i}n_{-i}+\mathbf{S_{+i}}\cdot \mathbf{S_{-i}}\right)
  \label{eq:trivial_interaction}
\end{align}

The first  and second terms are the   on-site  and nearest neighbor  repulsions respectively. The third term   is an inter-site Hund's interaction which preserves the $U(4)$ symmetry (as do the first two terms). The last two terms however break $U(4)$. The term proportional to $J_H'$  is the $U(4)$ symmetry breaking part of the nearest neighbor Hund's coupling (it breaks $U(4)$ down to $U(2)_+\times U(2)_-$. Finally the last term (proportional to $J_H$  is an on-site inter-valley Hund's coupling term which breaks $U(4)$ down to $U(1)_c\times U(1)_v\times SU(2)_s$ (upto  modding by a discrete $Z_2$ group) . Here $U(1)_c$ corresponds to the total charge conservation and $U(1)_v$ corresponds to the valley charge conservation. $SU(2))s$ is the spin rotation.  
In Table.~\ref{table;interaction_parameters} we list estimates of the parameters that enter the interaction Hamiltonian. We note that the dominant part of the interaction is given by the first 3 terms that preserve the $U(4)$ symmetry. Thus 
to leading order we can only consider the $SU(4)$ symmetric part in the interaction and view  the Hund's coupling $J'_H, J_H$ as small perturbations.

\begin{table}[h]
\centering
\begin{tabular}{c|c|c|c|c}

\hline
$U$ &$g_1 U$ & $2 g_h U$ & $J'_H$ & $J_H$\\ 
\hline
$25$&$10$& $0.4$ & $0.05$ &$0.136$\\
\hline
\end{tabular}
\caption{Parameters of interaction terms  in units of meV for $\Delta_V=-30$ meV. To estimate these parameters, we use a screened Coulomb interaction $V(\mathbf q)=\frac{e^2}{2 \xi_0 \kappa} \frac{1}{q}(1-e^{- q r_0})$ with $\kappa=8$ and  screening length $r_0=5 a_M\approx 75$ nm. $g_1\approx 0.4$ and $g_h\approx 0.008$ are estimated from Wannier orbital calculations explained in Appendix.~\ref{appendix:lattice_models}. The dependence of the interaction parameters on $\Delta_V$ is weak.}
\label{table;interaction_parameters}
\end{table}

Eq.~\ref{eq:kinetic_trivial} and Eq.~\ref{eq:trivial_interaction} give the   lattice model for $\Delta_V<0$. The dominant terms correspond to  a spin-valley extended Hubbard model on a triangular lattice. The most significant  $U(4)$ symmetry breaking is from the valley-contrasting flux in the hopping term. The interaction term is dominated by the on-site and nearest-neighbor Hubbard repulsion, which is guaranteed to be $SU(4)$ symmetric. However, there is also  a small ferromagnetic Hund's coupling term. Such a term  plays an important role in the spin physics of the Mott insulator though its value is only $2\%$ of the Hubbard $U$.  The lattice model has an approximate $U(2)_+\times U(2)_-$ symmetry, which is further broken down to $U(1)_c\times U(1)_v\times SU(2)_s$ by the on-site inter-valley Hund's coupling $J_H$ term.    

The inter-site Hund's coupling, like all the other interactions, emerge from projection of the Coulomb interaction.   Why does the pure density-density interaction give rise, after projection, to such a Hund's interaction?  
The reason is that the microscopic density operator has a complicated form in terms of the lattice operators: $\rho^{phy}(\mathbf x)\sim c^\dagger_{i;a\sigma}c_{i;a\sigma}+ a (c^\dagger_{i;+\sigma}c_{i;-\sigma}e^{-i 2\mathbf K_o \cdot \mathbf{x_i}}+h.c.)+b_{ij} c^\dagger_{i;a\sigma}c_{j;a\sigma}$ with $a,b_{ij}$ small but generically not zero. The $a$ term gives the on-site inter-valley Hund's coupling $J_H$ and the $b$ term gives the inter-site Hund's coupling $2g_hU$ and $J'_H$ terms. The $a$ term originates from the fact that the inter-valley bilinear $c^\dagger_{+}c_{-}$ gives an oscillating density wave with momentum $2\mathbf{K_o}$, where $\mathbf{K_o}$ is the large momentum in the original Brillouin Zone of a pure graphene layer.  The $b$ term comes from the fact that two nearest neighbor Wannier orbitals  $\langle ij \rangle$  are not tightly confined and their electron densities overlap\cite{po2018origin}.  As is well-known the Wannier orbital is gauge dependent and a natural question to ask is  if we  can  choose a good gauge to make  these orbitals  sufficiently tightly confined that   $b \approx 0$.   The answer is no:  the reason is   that local regions of the the valence band have non-zero Berry curvature (though there is no net Chern number). Such a non-zero Berry curvature is lost in the above one-orbital lattice model. The cost of this loss is that the microscopic density operator can not be purely on-site. In momentum space, $\rho(\mathbf q)\sim  \sum_{\mathbf k}\lambda_a(\mathbf k,\mathbf q)c^\dagger_{a;\mathbf{k+q}}c_{a;\mathbf{k}}$. The form factor $\lambda_a(\mathbf k,\mathbf q)\sim |F(\mathbf k)|e^{i \mathbf{A}(\mathbf k)\cdot \mathbf q}$ at small $\mathbf q$, where $\mathbf{A}(\mathbf k)$ is the Berry connection. Due to the non-zero Berry curvature, the form factor $\lambda_a(\mathbf k,\mathbf q)$ can not be equal to $1$ in any gauge.  Thus the  density operator can not be written as $\rho(\mathbf q)=\sum_{\mathbf k}c^\dagger_{a;\mathbf{k+q}}c_{a;\mathbf k}$ in any gauge. As a consequence, in the lattice model (for any gauge choice), the microscopic density operator can not be pure on-site, and will include   inter-site hopping terms. The original pure density-density interaction will then lead to density-density, density-hopping, hopping-hopping interaction in the lattice models.  As explained in the Appendix.~\ref{appendix:lattice_models}, there are several terms generated, like correlated hopping and pair hopping terms.  Of these the only term that does not  involve  double occupancy (which is suppressed by the Hubbard $U$)  is the inter-site Hund's coupling term $2g_h U$.

\subsection{Response to Magnetic Field: Valley Zeemann Coupling}
 Not only does  the one-orbital lattice model lose the information of the Berry curvature of the Bloch states, it also loses information on the   orbital  magnetic  moment.   It is well established that Bloch states have an orbital magnetic moment $m(\mathbf k)$ in the $z$ direction\cite{xiao2010berry}. A large $g$ factor for valley orbital magnetic moment has been proposed theoretically and verified experimentally in graphene systems\cite{xiao2007valley,koshino2011chiral,ju2017tunable}. A recent experiment sees evidence of a very large $g$ factor(of the order of hundreds) for valley orbital magnetic moment in monolayer graphene/h-BN system\cite{komatsu2018observation}. Motivated by these previous results, we study the possibility of a large valley orbital magnetic moment in the TG/h-BN system within the continuum model.
 
The corresponding g factor $g(\mathbf k)=m(\mathbf k)\frac{4 m_e}{\hbar^2}$ is

\begin{equation}
  g(\mathbf k)=-\frac{4 m_e}{\hbar^2} Im \sum_{n'\neq n}  \frac{\braket{n|\partial_{k_x}H|n'}\braket{n'|\partial_{k_y}H|n}}{\xi_n(\mathbf k)-\xi_{n'}(\mathbf k)}
  \label{eq:valley_g_factor}
\end{equation}
where we suppressed the valley index $a=\pm$ in $H_a(\mathbf k)$ and $\ket{n_a(\mathbf k)}$.  $n$ is the valence band and  $n'\neq n$ labels  the other eigenstates of $H(\mathbf k)$. 

Time reversal guarantees that $g_+(\mathbf k)=-g_-(-\mathbf k)$.  An   external  out-of-plane magnetic field $B$  couples linearly to this orbital moment:
\begin{equation}
  H_B=-B \sum_{\mathbf k}\left(g_+(\mathbf k)  c^\dagger_+(\mathbf k)c_+(\mathbf k)+g_{-}(\mathbf k)c^\dagger_{-}(\mathbf k)c_{-}(\mathbf k)\right)
\end{equation}

We calculated $g_a(\mathbf k)$ following Eq.~\ref{eq:valley_g_factor} within the continuum model. Generically $g_a(\mathbf k)$ has a strong dependence on momentum $\mathbf k$. Its behavior for $\Delta_V<0$ and $\Delta_V>0$  are qualitatively different. The modified band structures  that include this orbital  magnetic field are presented in Appendix.~\ref{appendix:band_structures}. 

\begin{figure}
\centering
\includegraphics[width=0.45\textwidth]{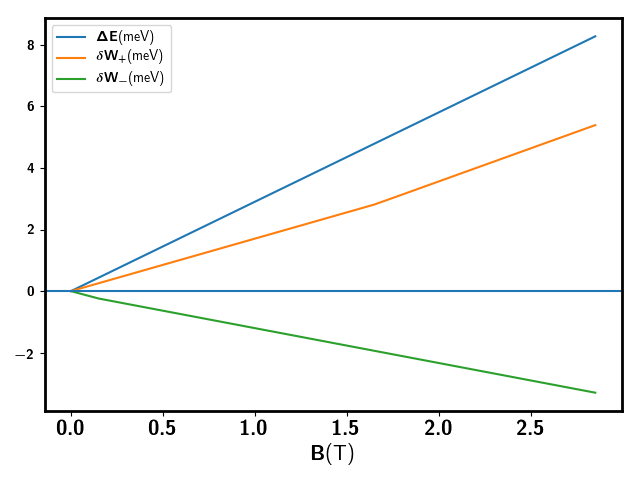}
\caption{Response to out of plane magnetic field $B$ from the valley Zeeman coupling at $\Delta_V=-25$ meV. $\Delta E=\bar{E}_+-\bar{E}_-$ is the splitting of the average energy of the valley $+$ and the valley $-$. $\delta W_a$ is the change of the bandwidth for valley $a$. A small magnetic field $B=1$ T split the average energy for two valleys by about $3$ meV. Meanwhile the bandwidth of one valley is increased by around $1.5$ meV while the bandwidth of the other valley is reduced by around $1.5$ meV.} 
\label{fig:response_to_magnetic_field}
\end{figure}

For $\Delta_V<0$, $g_{+}(\mathbf k)<0$ and $g_{-}(\mathbf k)>0$ for every $\mathbf k$. Therefore effectively we have a valley Zeeman coupling. The averaged $g$ factor is $\bar g\approx 54$, much larger than the $g=2$ for spin. Therefore for $\Delta_V<0$, the most dominant effect of a small out-of-plane magnetic field is the splitting of valley energy, rather than the familiar spin Zeeman effect. In addition to the splitting of the average energy of the two valleys, the out-of-plane magnetic field also increases the bandwidth of one valley while reducing the bandwidth of the other valley, as shown in Fig.~\ref{fig:response_to_magnetic_field}.

 At small $B<1$ T , this valley Zeeman coupling term can be used to polarize the valley in the Mott insulating regime.   A larger $B \sim 3$ T can greatly increase the total bandwidth of the two valleys, which could destroy the Mott insulating phases. $B=3$ T gives a flux per moir\'e unit cell $\Phi_B \approx 0.12\frac{h}{e}$. The system then crosses over to the Hofstadter butterfly region.

\section{Strong Mott Insulators}
We now discuss   the experimentally observed insulating states\cite{chen2018gate} at filling $\nu_T=1$ and $\nu_T=2$ for  $\Delta_V<0$  using the   model described in Section.~\ref{section:lattice_model_trivial}.  For the time being we only focus on the strong coupling limit $U>>t$. In this case charge is frozen and the low energy physics is governed by an effective spin-valley model.
At each site, we define the spin operator $\mathbf{S}=\frac{1}{2}c^\dagger_{a\sigma_1}\vec{\sigma}_{\sigma_1\sigma_2}c_{a\sigma_2}$ and  the valley operator $\mathbf{S}=\frac{1}{2}c^\dagger_{a_1\sigma}\vec{\tau}_{a_1a_2}c_{a_2\sigma}$. Here $\vec{\sigma}$ and $\vec{\tau}$ are Pauli matrices for the spin and the valley respectively\footnote{We have assumed Einstein summation convention.}.

Using  the standard $\frac{t}{U}$ expansion (see Appendix.~\ref{appendix:spin-valley model}) we find the spin-valley model:
\begin{align}
	H_S&=\frac{J_1}{8}\sum_{\langle ij \rangle}(1+\mathbf{\tau_i}\cdot \mathbf{\tau_j})(1+\mathbf{\sigma_i}\cdot \mathbf{\sigma_j})\notag\\
  &+\frac{J_2}{8}\sum_{\langle \langle ij \rangle \rangle}(1+\mathbf{\tau_i}\cdot \mathbf{\tau_j})(1+\mathbf{\sigma_i}\cdot \mathbf{\sigma_j})\notag\\
&+\frac{1}{8}\sum_{\langle ij \rangle} J^1_{p;ij}(\tau^x_i\tau^x_j+\tau^y_i\tau^y_j)(1+\mathbf{\sigma_i}\cdot \mathbf{\sigma_j})\notag\\
&+ \frac{1}{8} \sum_{\langle ij \rangle} J^2_{p;ij} (\tau^x_i\tau^y_j-\tau^y_i\tau^x_j)(1+\mathbf{\sigma_i}\cdot \mathbf{\sigma_j})\notag\\
&+O(\frac{t^3}{U^2})
  \label{eq:J1_J2_spin_model}
\end{align}
where $J_1=-2g_{h}U+\frac{4t_1^2}{\tilde U}$ with $\tilde U=(1-g_1)U (=0.6 U$ using the estimate in Table \ref{table;interaction_parameters}) and $J_2=\frac{4t_2^2}{U}$. $J_1$ has two contributions: a ferromagnetic part from the Hund's coupling and an anti-ferromagnetic part from the standard super-exchange. Here $\tau^\mu_i \sigma^\nu_i$ should be understood as tensor product and is the abbreviation of the bilinear term $c^\dagger_{i;a_1\sigma_1}\tau^\mu_{a_1 a_2}\sigma^\nu_{\sigma_1\sigma_2}c_{i;a_2\sigma_2}$. At $\nu_T=1$, $\tau_i$ and $\sigma_i$ are simply the corresponding valley and spin operator. At $\nu_T=2$, $\sum_{a\sigma}c^\dagger_{i;a\sigma}c_{i;a\sigma}=2$ and  the corresponding spin or valley operator at each site is a $4\times 4$ matrix, which can be generated from the above bilinear terms of the fermionic operator. The factor $\frac{1}{8}$ is added to make the $J$ consistent with the traditional convention in the spin $\frac{1}{2}$ model once valley is polarized.

$J^1_{p;ij}$ and $J^2_{p;ij}$ are the $SU(4)$ symmetry breaking terms, mainly originating from super-exchange term involving opposite valleys. The valley-contrasting phase in the hopping is inherited in this term. We have $J^1_{p;ij}=(J_1+2g_h U)(\cos 2\varphi_{ij}-1)+J'_H$ and $J^2_{p;ij}=(J_1+2 g_hU)\sin 2\varphi_{ij}$. The magnitude $|\varphi_{ij}|=\frac{|\Phi|}{3}$.  Here $\varphi_{ij}$ is the phase of the hopping for the valley $+$ of the bond $\langle ij \rangle $. $J'_H\approx 0.05$ meV  is from the  $SU(4)$ breaking part of the Hund's coupling and can be neglected.

In the above we ignore $t_3$ and $t_4$ for simplicity. One can easily add $J_3=\frac{4t_3^2}{U}$ and $J_4=\frac{4t_4^2}{U}$ terms. For the fourth neighbor coupling, the $SU(4)$ breaking term from the valley-contrasting hopping phase should also be considered because $t_4$ has a large phase.

Even at  second order of $\frac{t}{U}$ expansion, we  need to keep four parameters for the spin-valley model: $J_1$, $J_2$, $\Phi$ and $2g_h U$. Ferromagnetic Hund's coupling $2g_h U\approx 0.4$ meV is even larger than $J_1$ and can not be ignored. These parameters can be tuned by $\Delta_V$ and a rich phase diagram may be accessible in the experiment.  For $\Phi\sim 0.5 \pi - 2\pi$, $J^1_{p;ij}$ and $J^2_{p;ij}$ are generically of the same order of $J_1$. Therefore the $SU(4)$ symmetry is strongly broken to $SU(2)_{+}\times SU(2)_{-}\times U(1)_v$\footnote{Stictly speaking we need to further module some discrete symmetries}. For $\nu_T=2$, we also need to add the $J_H$ term in Eq.~\ref{eq:trivial_interaction} which further breaks down the symmetry to $U(1)_v\times SU(2)_s$.

\begin{figure}[ht]
\centering
    \includegraphics[width=0.45\textwidth]{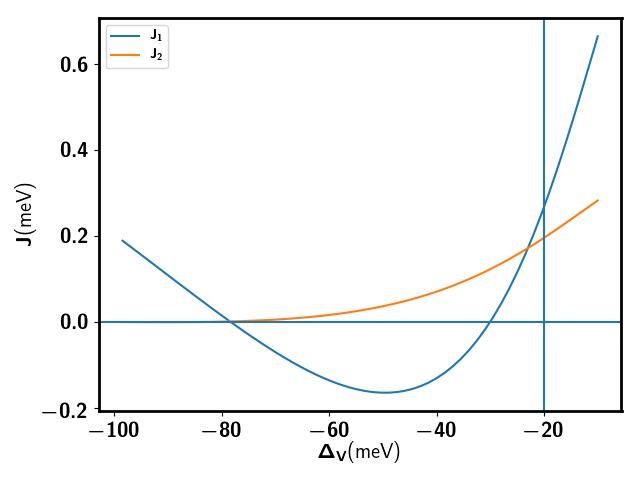}
  \caption{$J_1-J_2$ parameters with $\Delta_V$. We fix $U=25$ meV, $g_1=0.4$ and $g_h=0.008$ in the calculation. The vertical line is the value of $\Delta_V$ for which the bandwidth $W=U$. Deep inside the Mott insulating phase, $J_1$ is ferromagnetic from the Hund's coupling. In the intermediate regime, both $J_1$ and $J_2$ are antiferromagnetic.}
  \label{fig:lattice_parameters}
\end{figure}

A plot of $J_1-J_2$ with $\Delta_V$ is shown in Fig.~\ref{fig:lattice_parameters}. $J_1$ can be tuned to be either ferromagnetic or antiferromagnetic.   Though we have presented estimates of the parameters 
$J_1$, $J_2$, $\Phi$ and $2g_h U$, their precise  quantitative value  are sensitive to assumptions used in the band structure calculation\footnote{Even the sign of $J_1$ is sensitive to $g_h$. If $g_h$ is increased by a factor of $2$, $J_1$  will be ferromagnetic in the whole region of $U>W$.}. It is useful therefore to view them as phenomenological parameters  and discuss the general phase diagram of the model in Eqn. \ref{eq:J1_J2_spin_model}.

In the following two subsections we discuss the possible states for $J_1<0$ and $J_1>0$ region separately.

\subsection{Ferromagnetic Region}
 In the strict  limit $\frac{t}{U}\rightarrow 0$, the Hund's coupling  dominates over the other terms.  Then  $J_1<0$ and $J_2\sim J^1_{p;ij}\sim J^2_{p;ij}\sim 0$. The Mott insulator should thus be a spin-valley ferromagnetic state.

For $\nu_T=1$, the ground state should be ferromagnetic. The spin is polarized to any direction because of the $SO(3)$ spin rotation symmetry. For the valley, we need small anisotropic terms to decide whether $\tau_x$ or $\tau_z$ order is favored. The small $SU(4)$-breaking  Hund's coupling $J'_H$ ($\sim 0.05$ meV) term in Eq.~\ref{eq:trivial_interaction}  favors   $\tau_z$ valley polarization. But the anisotropy inherited from the valley-contrasting hopping term in $J^1_{p;ij}$  of in Eq.~\ref{eq:J1_J2_spin_model} favors the $\tau_x$ polarization.  Therefore interaction term and kinetic term compete with each other. At the flat band limit we always have the $\tau_z$ valley polarization.  At any non-zero temperature $T$, the spin ferro-magnetism  will be disordered immediately because of the Mermin-Wagner theorem.  However, valley polarization only breaks a discrete time reversal symmetry and will therefore be stable upto a finite temperature continuous transition in the Ising universality class.  The spontaneous breaking of time reversal at small non-zero $T$ may give an exponentially suppressed but non-zero Hall conductivity.  Such a valley polarization may also be detectable via the magneto-optical Kerr effect, as demonstrated in Ref.\onlinecite{huang2017layer} for spin ferromagnetism.  As the out-of-plane magnetic moment from the valley is $20$ times larger than spin, this effect should be more significant for the valley polarized state. Once $t/U$ is increased, there can be a phase transition to an Inter-valley-coherent (IVC) order ($\tau_x$ polarization).  The IVC order does not break the time reversal symmetry. As it breaks the $U(1)_v$ symmetry, there can be a Berezinskii-Kosterlitz-Thouless transition (BKT transition) at finite temperature.

For $\nu_T=2$, just from the $SU(2)_+\times SU(2)_-\times U(1)_v$ symmetric interaction in Eq.~\ref{eq:J1_J2_spin_model} there are several degenerate states.   The true ground state will be selected from these by small anisotropies. The onsite inter-valley Hund's coupling $J_H$ in Eq.~\ref{eq:trivial_interaction} will select the spin polarized, valley singlet state as the ground state. Such a spin ferromagnetic state cannot have true long range order  at any non-zero $T$.  

In summary, for $\frac{t}{U}\rightarrow 0$ limit, the ground states for both $\nu_T=1$ and $\nu_T=2$ are ferromagnetic.  There should be a finite temperature transition corresponding to the valley polarization for $\nu_T=1$ and no transition for $\nu_T=2$. We emphasize that the destruction of the spin ferromagnetism  at finite temperature does not close the charge gap, which is at order $U$ and is thus much larger than the ferromagnetic scale $J_1\sim 0.01 U$.

\subsection{Antiferromagnetic Region}
With increasing $\frac{t}{U}$, we enter a regime dominated by the the antiferromagnetic super-exchange: $J_1, J_2 >0$.  The  frustrated triangular geometry and the larger number of degrees of freedom\footnote{In the limit where we only keep $J_{1,2}$ we get an $SU(4)$ antiferromagnet with spins in either the fundamental representation (at $\nu_T = 1$) or in the 6-dimensional representation (at $\nu_T = 2$) of $SU(4)$. Such  models, even when nearest neighbor, are more likely to be in non-magnetic ground states than their $SU(2)$ versions.  } than the standard spin-$1/2$ model both enhance the effect of quantum fluctuations. Density Matrix Renormalization Group (DMRG)  calculations  of Eq.~\ref{eq:J1_J2_spin_model}  may be able to map the phase diagram. Here we restrict ourselves to brief comments about special cases where we can relate the model to others studied in the literature. 
 At  $\nu_T=1$, because of  of the large valley Zeeman effect, a small out of plane magnetic field (of order  $\approx 0.2$ T) can already give an energy splitting larger than $J_1$ and $J_2$. Then the valley is frozen into a polarized state, and the effective model becomes the standard Heisenberg $J_1-J_2$ spin $\frac{1}{2}$ model. This model  is already well studied\cite{zhu2015spin,hu2015competing,gong2017global}.  At small $\frac{J_2}{J_1}$, the ground state is the well known $120^\circ$ magnetically ordered state. At large $\frac{J_2}{J_1}$ ratio the ground state is   a stripe  antiferromagnet. In the intermediate region, a spin liquid phase is suggested from  DMRG calculations\cite{zhu2015spin,hu2015competing} though precisely which kind of  spin liquid is not clear. Candidates are a chiral spin liquid or a $U(1)$ Dirac spin liquid. 
 Another special case is to apply a large in-plane magnetic field to polarize the spin. We expect then that the  remaining valley degree of freedom   forms  a $120^\circ$ order at small $\frac{J_2}{J_1}$.

\section{Weak  Mott insulators: possibility of a continuous Mott Transition}
We now discuss the region close to the Mott metal-insulator transition for $\Delta_V<0$.    In this region the spin-valley model derived in the previous section  will not be adequate to discuss the Mott insulator. We could keep higher order terms in the $\frac{t}{U}$ expansion  which will include multi-site ring exchange processes\cite{lesik2005mott}. Alternately  the physics  (even in the insulating side) may be directly discussed within the framework of the original Hubbard model. 

The Mott transition is of course most central to the study of correlated electron systems, and there is a vast literature\cite{imada1998metal}. 
It has long been appreciated that there are many distinct routes by which a metal may evolve into a Mott insulator at zero temperature. 
 A common fate (realized in many experimental systems) is that the transition occurs between the paramagnetic metal and a magnetic insulator and is first order.  Such a route can potentially be avoided in frustrated low dimensional lattices (as pertinent to the present paper).  A different route\cite{krish1990mott}, suggested by a simple Hartree-Fock theory for an antiferromagnetic order parameter\footnote{In the following we will use the term `magnetic' to  denote ordering in the spin-valley space.},  is that the paramagnetic metal first undergoes a magnetic ordering transition into a magnetic metal. Eventually there is a second transition where the magnetic metal becomes a magnetic insulatior.   A third fascinating alternative is that there is a continuous quantum critical Mott transition. A theory for such a continuous Mott transition\cite{senthil2008theory} exists when the Mott insulator is a quantum spin liquid with a neutral spinon  Fermi surface coupled to a $U(1)$ gauge field.  Such a continuous Mott transition may be relevant to experiments\cite{kanoda2005mott,kanoda2018mott} on quasi-two dimensional organics.  It is currently not at all clear if other kinds of Mott insulators admit continuous zero temperature quantum phase transitions into the paramagnetic metal. 

\begin{figure}[ht]
\centering
    \includegraphics[width=0.45\textwidth]{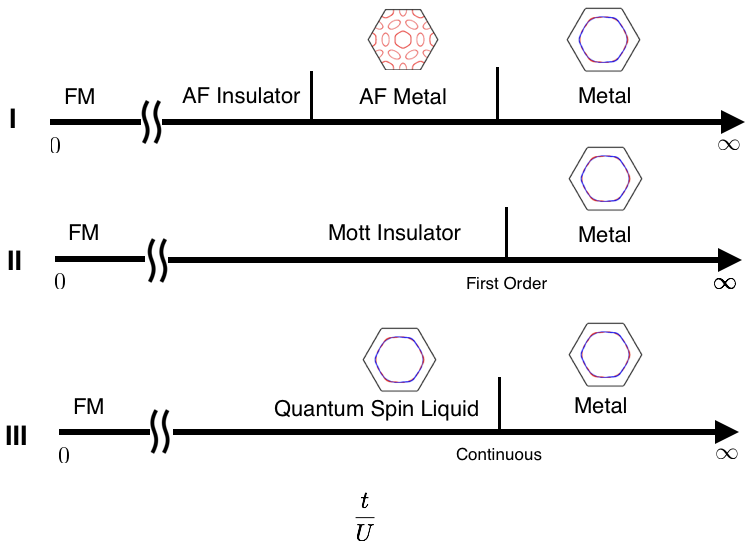}
  \caption{Three possible phase diagrams tuned by $\frac{t}{U}$ at $\nu_T=2$. In (I) AF Metal means metal coexisting with antiferromagnetic order. AF insulator is a Mott insulator with antiferromagnetic order (The most likely candidate is the $120^\circ$ valley order). In (II) the Mott insulator may be antiferromagnetic  or may be a quantum spin liquid. In (III) the specific quantum spin liquid we consider has a  spinon Fermi surface coupled to a $U(1)$ gauge field.   At $\frac{t}{U}=0$, the ground state is a ferromagnet because of inter-site Hund's coupling.  We also show the plots of the Fermi surfaces. The Fermi surfaces are calculated at $\nu_T=2$ using $t_1,t_2,t_3,t_4$ for $\Delta_V=-20$ meV. For the AF metal, we use the $120^\circ$ inter-valley order with the order parameter $M= 2 |t_1|$. The Fermi surface area should decrease continuously in the AF metal region as $\frac{M}{|t_1|}$ increases.}
  \label{fig:metal_insulator_trivial}
\end{figure}

 The three possible evolutions discussed above from metal to Mott insulator are illustrated in Fig.~\ref{fig:metal_insulator_trivial}.

The  TG/h-BN (and other graphene moire systems)  offers a  tremendous opportunity to explore the band-width controlled Mott   transition  in a frustrated two dimensional lattice.  There is a large body of very interesting prior work (see for instance Refs.\onlinecite{ngkanodarmp,kanoda2005mott,kanoda2015mott,kanoda2018mott,kanoda2011arcmp} ) on  quasi-two dimensional organic salts (also on triangular lattices) which has probed the Mott transition with pressure as a tuning parameter at low temperature.   Compared to the organics, the graphene system has  the advantage that the electric control of bandwidth should make it a lot easier to tune through the Mott transition at low temperature and study it in exquisite detail. 

With this in mind below we propose concrete (and we believe, feasible, in TG/h-BN) experiments that  distinguish   these various routes to the Mott transition.

\subsection{ `Magnetic' metal as an intermediate phase}
We first consider the situation where the evolution from the metal to an antiferromagnetic (in spin-valley space)  Mott insulator occurs in two stages. First there is a phase transition inside the metallic phase where the entiferromagnetic order onsets  leading to a modification of the unit cell. This reconstructs the Fermi surface. With increasing amplitude of the antiferromagnetic  order parameter, the Fermi surfaces  will shrink and there will be a further transition to an antiferromagnetic insulator. This is the natural result of a Hartree-Fock treatment of the interactions. 
In the TG/h-BN context, such a symmetry breaking  is suggested to arise from the nesting of the Fermi surfaces for $\nu_T=2$ by Ref.~\onlinecite{zhu2018antiferro}. Nesting driven theories have also been proposed for the twisted bilayer graphene system\cite{isobe2018superconductivity,you2018superconductivity}.

 \begin{figure}[ht]
\centering
\includegraphics[width=0.45\textwidth]{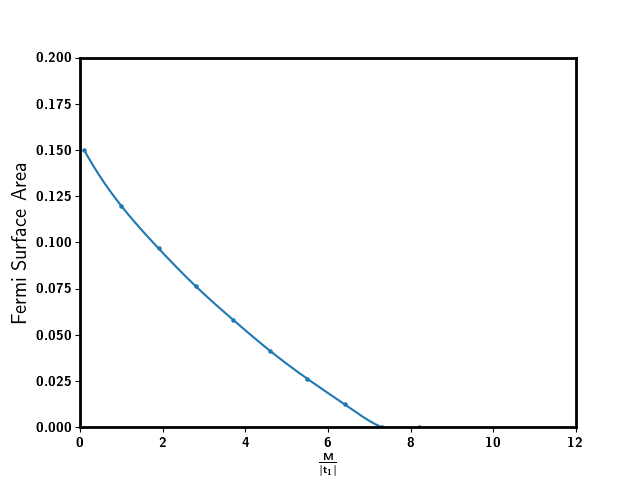}
\caption{The change of Fermi surface area(in units of the area of the MBZ) with order parameter $M$ for the $120^\circ$ inter-valley order: $H_M=-M \sum_{\mathbf x}c^\dagger_{\mathbf x}(\cos(\mathbf{Q}\cdot \mathbf{x})\tau_x+\sin(\mathbf{Q}\cdot \mathbf{x})\tau_y)c_{\mathbf x}$ with $\mathbf{Q}=(\frac{4\pi}{3},0)$. We use $t_1=2.14 e^{i 0.141 \pi}$ meV and $t_2=-1.372$ meV for $\Delta_V=-20$ meV.  There are several Fermi surfaces and we only count the hole pocket at $\Gamma$ point. At $M\rightarrow 0$, magnetic breakdown effect should give a quantum oscillation  frequency corresponding to the original Fermi surface area equal to $0.5$, which is not captured by our calculation here. After adding a non-zero $t_2$, Fermi surfaces can not be fully gapped out until $M=8|t_1|$. } 
\label{fig:AF_metal_FS_area}
\end{figure} 

A clear experimental probe  of this scenario is to study Shubnikov-DeHaas (SdH) oscillations in the resistivity in a  perpendicular magnetic field.  Through out the paramagnetic metal phase the Fermi surface area, and hence the SdH frequency,  is fixed to be a constant by Luttinger's theorem.  In the antiferromagnetic metal, the reconstruction of the Fermi surface will change the SdH frequencies. On approaching the insulator these frequencies will decrease (possibly all the way to zero if the transition from the antiferromagnetic metal to antiferromagnetic insulator is continuous). Thus in this scenario there will be a change in the SdH frequencies {\em before} the metal becomes an insulator similar to Fig.~\ref{fig:AF_metal_FS_area}. We caution that the SdH experiments should be performed in {\em low} perpendicular magnetic field so that they are a soft probe of the Fermi surface of the metal. At larger fields we will enter the quantum Hall regime and the oscillations may not directly reveal the Fermi surface structure of the zero field metal. 

Let us briefly further comment on this simple Hartree-Fock  scenario.  In the strong Mott insulating region, the  system may possibly be in a spin-valley ordered  antiferromagnetic phase. However the mechanism for such ordering is different in the metal  where it may be driven by an approximate  nesting of the Fermi surface. Ref.~\onlinecite{zhu2018antiferro} suggested such a nesting driven mechanism for $\nu_T=2$ by using a nearest neighbor tight binding model with valley contrasting flux $\Phi=\frac{\pi}{2}$. However, according to our calculation in Fig.~\ref{fig:hopping_flux}, the flux $\Phi$ is generically not equal to $\frac{\pi}{2}$ and $t_2,t_3,t_4$ are also necessary to reproduce the band structures. One natural question is  whether this nesting of Fermi surfaces at $\nu_T=2$ is fine tuned or not. To test the robustness of the nesting properties of the Fermi surfaces, we calculated the Density of States(DoS) at $\Delta_V=-5,-10,-15,-20,-25,-30,-40$ meV using the continuum model with a $300\times 300$ mesh-grid in momentum space. The Van-Hove singularity in our model is away from the Fermi level at both $\nu_T=1$ and $\nu_T=2$ as shown in Fig.~\ref{fig:dos}. From the Fermi surface plots in Appendix.~\ref{appendix:band_structures} one can also see that there is no nesting instability in the particle-hole channel.  Thus it is not obvious that the Hartree-Fock scenario is realized in the  experimental system.  We will therefore consider also other scenarios  for the evolution from metal to insulator.


\subsection{First order Mott transition}  
A common possibility is that there is a first order transition between the paramagnetic metal and a Mott insulator.  This may happen irrespective of the detailed description of the insulator (antiferromagnetic or quantum spin liquid). 
 In this scenario the Fermi surface area seen in quantum oscillations should be constant in the metallic region.  The first order transition  will be  accompanied by    hysteresis  when  $D$ is cycled through the metal-insulator transition.

 Further a $T = 0$ first order transition will continue to $T \neq 0$ (till a critical end-point in the Ising universality class) as a sharp transition. Hysteresis will be observed on crossing this finite $T$ phase boundary. If such a first order transition is indeed seen the shape of the transition line in the $T-D$ plane may provide some clues\footnote{Specifically, through the Claussius-Clapeyron relation, the metal-insulator phase boundary will tilt toward the insulator or metal depending on which state has more entropy at a given low $T$. An antiferromagnetic insulator will  at low-$T$ have lower entropy than the metal while some spin liquid insulators have higher entropy than a metal. } about the nature of the Mott insulator.

\begin{figure}[ht]
\centering
    \includegraphics[width=0.45\textwidth]{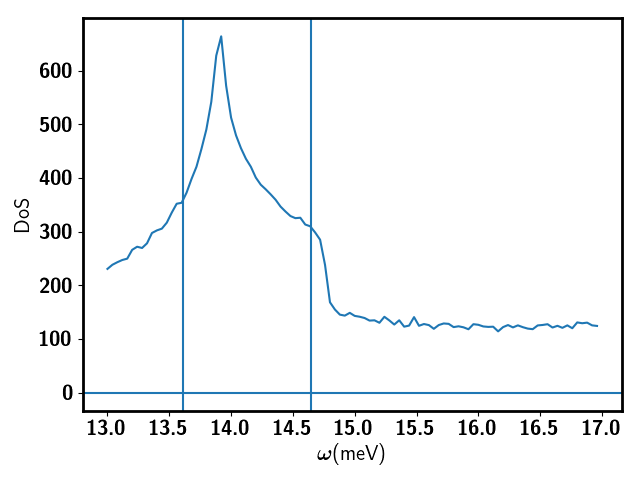}
  \caption{Density of state at $\Delta_V=-25$ meV. Two vertical lines correspond to $\nu_T=1$ and $\nu_T=2$. The Van Hove singularity is away from both $\nu_T=1$ and $\nu_T=2$. This is true for other values of $D$ in the region $-40<\Delta_V<-5$ meV. The closest distance to $\nu_T=1$ for the Van-Hove singularity is still at least $10\%$ doping away. The Van-Hove singularity is associated with a Lifshitz transition of the Fermi surfaces(See Appendix.~\ref{appendix:band_structures}). At exactly $\nu_T=1,2$, there is no obvious instability for the Fermi surfaces.}
  \label{fig:dos}
\end{figure}

 \subsection{Bandwidth Controlled Continuous Metal-Insulator Transition}

 It is hard to theoretically rule out either of the two scenarios described above. However for the simpler problem of the spin-$1/2$ triangular lattice Hubbard model, it seems (from numerical studies\cite{lesik2005mott,imada2002qsl,mila2010qsl,sheng2009spin,szasz2018observation}) that a quantum spin liquid state forms in the weak Mott insulating regime. Many existing numerical calculations\cite{lesik2005mott,imada2002qsl,mila2010qsl,sheng2009spin} as well as experiments\cite{kanoda2011arcmp,ngkanodarmp} on the organics are broadly consistent with this being a   spin liquid with a neutral Fermi surface.  A recent DMRG calculation\cite{szasz2018observation} however reports instead a gapped chiral spin liquid in the weak Mott region. The TG/h-BN system has more degrees of freedom (than the spin-$1/2$ Hubbard model) at each site which may make a spin liquid more likely in this regime. 
 
 A remarkable feature of the neutral Fermi surface state  is that it admits a continuous Mott transition to the metal. We turn therefore to how to  look for  this experimentally.

We first  review  a  (small generalization of a)   theory\cite{senthil2008theory}  for  the continuous Mott transition between a Fermi liquid metal  and a spin liquid Mott  insulator with a spinon Fermi surface   coupled to a $U(1)$ gauge field.   The theory should work for both $\nu_T=1$ and $\nu_T=2$. We use the  slave boson construction\cite{florens2004mott}: write $\psi_{a\sigma}({\mathbf x}) =b(\mf{x})  f_{a\sigma}({\mf x})$. Here  $b({\mf x})$ is a boson that carries the electric charge of the electron but not its spin/valley quantum numbers and $f_{a\sigma}({\mf x})$ (the spinon) is an electrically neutral  fermion that carries the spin/valley quantum number.  There is a constraint $n_b =n_f=n_\psi$ relating the number of $b, f$ and $\psi$ particles at each site of the lattice. Correspondingly  there is a $U(1)$ gauge redundancy $b({\mf x}) \rightarrow b({\mf x}) e^{i \alpha({\mf x})}$ and $f_{a\sigma}\rightarrow f_{a\sigma}e^{-i\alpha({\mf x})}$.  A reformulation of the original electronic problem in terms of the $(b, f)$ variables necessarily must include a dynamical $U(1)$ gauge field.  In the Fermi liquid phase the spinons form a Fermi surface while  $ \langle b \rangle \neq 0 $, {\em i.e}, the bosons are in a superfluid state.   Upon increasing interactions, a   Mott insulator will form.  Within this slave particle framework a natural Mott insulator is  obtained by letting $b$ form a bosonic Mott insulator (where $\langle b \rangle = 0$ while keeping the $f$-Fermi surface\cite{sslee2005mott}. The resulting state is a spin liquid Mott insulator. The Mott metal-insulator transition is then associated\cite{florens2004mott,senthil2008theory,mishmash2015continuous} with the superfluid- Mott transition of the boson $b$ in the presence of the spinon fermi surface and the $U(1)$ gauge field. As shown in Ref. \onlinecite{senthil2008theory} the resulting theory admits a continuous Mott transition which further is tractable.  We now highlight two predictions of this theory  for transport experiments that may be  directly feasible in TG/h-BN.

The first pertinent prediction is a universal jump\cite{senthil2008theory,krempa} by $R \frac{\hbar}{e^2}$of the residual resistivity as the Mott  critical point is approached from the metallic side\footnote{A simple explanation is from the Ioffe-Larkin rule which states that the physical resistivity $\rho = \rho_b + \rho_f$ where $\rho_{b,f}$ are the boson and $f$-fermion resistivities respectively. Across the Mott transition, $\rho_f$  evolves smoothy while $\rho_b$  goes from $0$ (in the metal) to a universal constant $= R \frac{\hbar}{e^2}$ (at the critical point) and eventually is $\infty$ (in the insulator ).   The universal resistivity jump follows}. Here $R$ is a universal number of $O(1)$. At a non-zero temperature the resistivity follows a useful scaling form described in Ref. \onlinecite{krempa}: 
 \be
 \label{rhoscal}
 \rho(T, \delta) - \rho_m = \frac{\hbar}{e^2} G\left(\frac{\delta^{z\nu}}{T}\right)
 \ee
with $z = 1$, and $\nu \approx 0.672$ in a clean sample. $\rho_m$ is the residual resistivity in the metal just before the Mott transition and $\delta$ is the parameter used to tune across the transition. For TG/h-BN this is accomplished very simply by the perpendicular displacement field. Thus the TG/h-BN system offers a promising platform to  access such a continuous Mott transition.  

A second prediction enables directly detecting the neutral Fermi surface, if it exists,  just on the insulating side of the Mott transition: such a neutral Fermi surface will lead to SdH oscillations\cite{motrunich2006orbital,chowdhury2018mixed,sodemann2018quantum}  in a weak Mott insulator.   Detailed expressions for the temperature dependence of such oscillations may be found in Ref. \onlinecite{sodemann2018quantum}.  The key point is that though the spinons  are electrically neutral, they couple to the internal $U(1)$ gauge field $\mathbf{a}$ which locks to an external field $\mathbf{A}$: $\mathbf{a}=\alpha \mathbf{A}$ with a factor $\alpha<1$. In the vicinity of Mott transition point, $\alpha$ will be of  order $1$. Therefore, the spinon fermi surface experiences an internal magnetic field $\mathbf{b}=\alpha \mathbf{B}$ and show quantum oscillation in the resistivity $\rho_f$. At finite temperature, $\rho_b$ is large but finite even inside the Mott insulator, and therefore $\rho =\rho_b+\rho_f$ should also show quantum oscillation with frequency enlarged by a factor of $\frac{1}{\alpha}$ compared to the Fermi liquid side. $\alpha$ should show dependence on voltage $D$ and also temperature (see Ref. \onlinecite{sodemann2018quantum}).  Due to the large valley Zeeman coupling, in practice, the oscillations may not have perfect periodicity in $\frac{1}{B}$. However, an oscillating response to $B$ inside a Mott insulator will be strong evidence of the existence of neutral Fermi surface and emergent gauge field.
Remarkably SdH  oscillations in electrical resistivity have been reported in a   recent experiment on a mixed valence insulator\cite{xiang2018quantum}.

Another more direct evidence for a spinon Fermi surface state is metallic thermal transport $\sigma_{thermal}\sim T$. Measurement of the thermal conductivity is hard, but may be possible in the future.

We emphasize that the only currently understood theory  for such a  continuous Mott transition  is when the insulator is a  $U(1)$ spin liquid with  an emergent neutral  fermi surface\cite{senthil2008theory}.  It is not known if there could be a direct continuous Mott transition between the paramagnetic metal and other kinds of Mott insulators (for instance an antiferromagnetic insulator or a chiral spin liquid). Such a continuous transition is exotic and will presumably involve a novel formulation.   In TG/h-BN if none of the signatures discussed above are seen it will provide experimental evidence for such an exotic continuous quantum phase transition.

\subsection{Doping Controlled Continuous Metal-Insulator Transition}

\begin{figure}[ht]
\centering
    \includegraphics[width=0.45\textwidth]{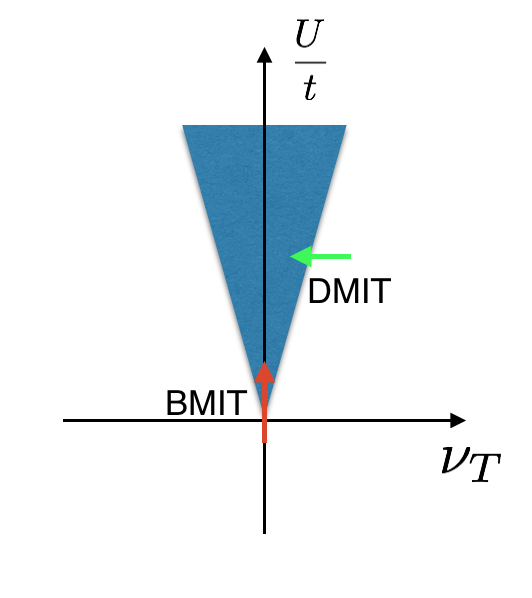}
  \caption{Illustration of Bandwidth controlled Metal-Insulator Transition (BMIT) and Doping controlled Metal-Insulator Transition (DMIT). The shaded region is the Mott insulator.}
  \label{fig:dmit}
\end{figure}
We now briefly address the Mott metal-insulator transition induced by doping away from commensurate filling.  We will restrict to a discussion of the possibility of a continuous Mott transition\footnote{Strictly speaking continuous Mott transitions are also possible out of paired spin liquid states. For instance if we dope a $Z_2$ spin liquid a natural outcome is a superconductor.  We then have a continuous Mott insulator -superconductor transition. } which is possible if the Mott insulator is in the quantum spin liquid with a spinon Fermi surface.  Theoretical descriptions of this Doping controlled Metal-Insulator Transition (DMIT)  may be found in Refs. \onlinecite{senthil2008theory,senthilleeifl2009,senthil2004weak}. Similar to our description of the Bandwidth controlled Metal-Insulator Transition (BMIT) in the previous subsection, we still use the slave boson theory: $c=bf$. In this case boson $b$ goes through a chemical potential tuned superfluid-Mott insulator transition. We focus here on the predictions for electrical transport.  From the Ioffe-Larkin rule  $\rho_c=\rho_b+\rho_f$.  In the clean limit  at a small but non-zero $T$ it is known\cite{senthil2004weak} that the bosons have a resistivity $\rho_b \sim \frac{1}{\log\frac{1}{T}}$ due to scattering from (Landau-damped) gauge fluctuations. The weak logarithmic dependence may not be visible, and hence we may roughly expect the residual resistivity to jump as the critical point is approached from the metallic side just like at the BMIT. 

Disorder effects will further affect the nature of the transition. First it is natural that at very low densities the dopants will be localized. The DMIT will then happen at a non-zero critical doping.  The bosons are expected to have a universal conductivity at this disordered critical point which is distinct from that in the BMIT case. Thus,  close to the critical point, we will once again have a universal jump of residual resistivity.   Finally we note that near the disordered critical point, scaling similar to Eqn. \ref{rhoscal} will hold but with different values for the exponents $z$ and $\nu$.  From the general result  $\nu\geq \frac{2}{d}=1$ ( where $d=2$ is the spatial dimension) for disordered critical points, and the expectation $z = 1$ in the presence of Coulomb interactions, we have  $z\nu \geq 1$ for the DMIT, larger than $z\nu\approx 0.672$ for the BMIT of a clean system.

This brief discussion was meant to motivate an experimental  study of the doping induced Mott transition in TG/h-BN. Interestingly the existing experimental data may already have evidence for a continuous doping controlled metal-insulator transition (DMIT) close to $\nu_T=2$. In the Fig.3(a) of Ref.~\onlinecite{chen2018gate}, there is a critical $V^c_t\approx -4.7$ V for $V_t$ which controls the total density (and also the bandwidth). Resistance $R$ increases with temperature $T$ when $V_t<V^c_t$  while when $V_t>V^c_t$ the resistance $R$ decreases with $T$. At exactly $V^c_t$ the resistance is finite (around $0.7 \frac{h}{e^2}$) and constant in the temperature region $1.5-40$ K. Here $1.5$ K is the lowest temperature reachable in the reported experiment in Ref.~\onlinecite{chen2018gate}. This suggests  a continuous metal-insulator transition.  As a further test , we suggest  measurements at lower temperature and   to  scale  the data according to Eqn. \ref{rhoscal} but with modified exponents as discussed above. It is also interesting to study the temperature dependence of the resistivity close to the critical point to search for non-Fermi-liquid behavior.  

Finally within the theory of Ref. \onlinecite{senthil2008theory} the quasiparticle effective mass in the metallic phase will diverge as $\frac{1}{\sqrt{\delta}}$ 
 (upto log corrections) where $\delta$ is the doping away from the Mott insulator. This strong divergence may be observable through SdH measurements. (In contrast at the BMIT a much weaker log divergence of the effective mass is predicted). 
 
\section{Comments for The $\Delta_V>0$ side \label{section:lattice_model_chern}}
 
When  $\Delta_V>0$, the valence bands of two valleys have non-zero Chern numbers $C=\pm 3$. Therefore it is not possible to construct localized Wannier orbitals for each valley separately.  Following a similar construction\cite{po2018origin} for twisted bilayer graphene,  we can  construct  a two orbital model on the triangular lattice (see Appendix.~\ref{appendix:Chern-lattice-models})  but with a non on-site implementation of the the valley charge operator ({\em i.e}, the valley charge operator is not a sum of on-site terms).  As a consequence, the interaction is in a complicated form, which makes an analytical treatment of the model very hard. Such a model may be useful for future numerical simulations.

\begin{figure}[ht]
\centering
\includegraphics[width=0.45\textwidth]{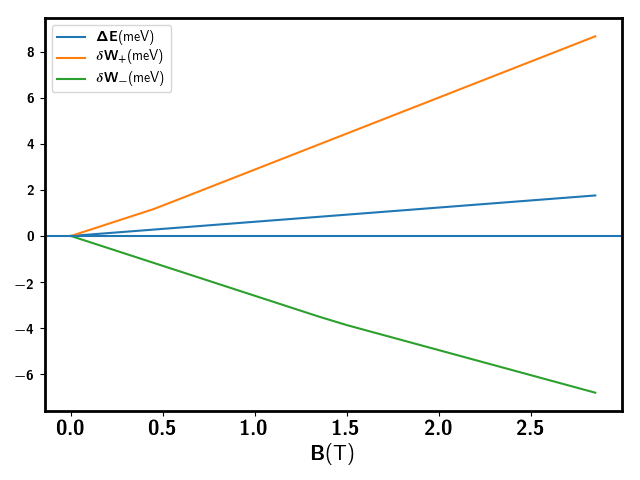}
\caption{Response to out of plane magnetic field $B$ from the valley Zeeman coupling at $\Delta_V=25$ meV. $\Delta E=\bar{E}_+-\bar{E}_-$ is the splitting of the average energy of the valley $+$ and the valley $-$. $\delta W_a$ is the change of the bandwidth for valley $a$.} 
\label{fig:response_to_magnetic_field_chern}
\end{figure}

Despite the complexity of the model, the $\Delta_V>0$ side can potentially realize interesting phases that show  the  Quantum Anomalous Hall effect (QAHE) and even the Fractional Quantum Anomalous Hall Effect (FQAHE) as proposed in our previous paper\cite{zhang2018moir}. Especially, similar to  quantum Hall ferromagnets, the $\nu_T=1$ insulator in the flat band limit should be a spin and valley polarized Chern insulator with Hall conductivity $\sigma_{xy}=3\frac{e^2}{h}$ even at zero magnetic field.  
One concern about the experimental realization of this QAHE state is that the energies of the two valley polarizations  are degenerate at zero magnetic field and  hence the system forms domains. However one can align the valley polarization by cooling  in an  out-of-plane magnetic field.  As shown in Fig.~\ref{fig:response_to_magnetic_field_chern}, there is also a valley Zeeman coupling  when $\Delta_V > 0$. The averaged $g$ factor is not so large as the $\Delta_V<0$ side because $g(\mathbf k)$ changes sign in the MBZ. However,  within our model, for a $z$ direction magnetic field with $1$ T, the band width of one valley becomes $6$ meV smaller than the other valley. Therefore, one valley polarization should be selected by a  magnetic field and the system will be in  the QAHE state. The total filling of the QAH insulator should also change with the magnetic field, leading to an insulating Landau fan: $\nu_T=1-3 |\Phi_U|$ where $\Phi_U$ is the uniform flux per moir\'e unit cell in units of $\frac{h}{e}$. For zero twist angle, $|\Phi_U|\approx 0.04$ for $B=1$ T.

The proposal of quantum Hall ferromagnetism in our previous paper \cite{zhang2018moir} assumes the flat band limit $\frac{W}{V}\rightarrow 0$. The possible phases at intermediate $\frac{W}{V}$ remain an open question, as does the nature of the evolution from the weak interacting metal.   A simple  possibility is that there is a  an intermediate ferromagnetic metallic phase which then gives way to the ferromagnetic insulator.  Clarifying this will require developing tools to deal with strong correlations in partially filled dispersing $\pm$ Chern bands  which we leave for the future.

\section{Conclusion}
In this paper we  discussed several aspects of the moir\'e superlattice system  in  ABC stacked trilayer graphene on hexagonal boron nitride where previous work has shown that an applied vertical electric field $D$ can tune both the bandwidth and the topology. Our focus in this paper was complementary to our earlier work which mainly discussed the phenomenology of the topologically non-trivial side ($\Delta_V > 0)$. Here we mainly discussed the other topologically trivial side ( $\Delta_V<0$). We explicitly constructed  a lattice extended   Hubbard model with $SU(4)$ degrees of freedom (but no $SU(4)$ symmetry).  We used this model as a framework to discuss possible Mott insulating states at   at total filling $\nu_T=1$ and $\nu_T=2$.    We also showed that due to a large valley Zeeman coupling a small perpendicular magnetic field may be a useful knob in this system.

We emphasized the opportunities provided by TG/h-BN (and other graphene moir\'e structures) to carefully experimentally study the bandwidth tuned Mott metal-insulator transition in a frustrated two dimensional lattice.  We showed how simple electrical transport experiments can distinguish many different routes to the Mott transition. Particularly exciting is the possibility that this system realizes a quantum spin liquid with a spinon Fermi surface in the vicinity of the Mott transition. Such a state  admits a direct continuous Mott transition to the Fermi liquid metal.  The transport experiments we describe  can specifically also probe this state and the continuous Mott transition.

Finally  when $\Delta_V>0$ and the bands  have Chern number  $C=\pm 3$, we constructed a lattice  two orbital model on the triangular lattice but with a non-local implementation of the valley charge operator (along the lines of the treatment of twisted bilayer graphene in Ref. \onlinecite{po2018origin}).   It remains to be seen whether this kind of model can be useful for a future attack on strongly correlated partially filled $\pm$ Chern bands.

\section{acknowledgement}
We thank Yuan Cao,  Debanjan Chowdhury, Mao Dan, Pablo Jarillo-Herrero, Adrian Po, Cecile Repellin, Ashvin Vishwanath, Feng Wang, Liujun Zou and Mike Zaletel  for many inspiring discussions. This work  was supported by NSF grant
DMR-1608505, and partially through a Simons Investigator
Award from the Simons Foundation to Senthil Todadri.

\bibliographystyle{apsrev4-1}
\bibliography{tg_bn}

\onecolumngrid
\appendix

\section{Band Structures\label{appendix:band_structures}}

First we give a brief introduction to the continuum model approach  used in Ref.~\onlinecite{zhang2018moir} and the current paper. If the two layers have slightly different lattice constants $a_1$ and $a_2$, or a small twiste angle $\theta$, then there is a moir\'e super lattice with  lattice constant $a_M\approx \frac{a}{\sqrt{\xi^2+\theta^2}}$ where $\xi=\frac{|a_1-a_2|}{a_2}$.  For TG/h-BN system, even if the twist angle $\theta=0$, there is still a moir\'e superlattice with $a_M\approx 58$ a, where $a\approx 0.246$ nm is the lattice constant for the graphene layer. Besides, we treat the two valleys separately. The two valleys are related by time reversal transformation. Therefore, we can do calculations for only one valley, for example, valley $+$.

First we ignore the h-BN layer. Then the ABC stacked trilayer graphene has cubic band touching at two momentum points $K_o$ and $K'_o$ in the original Brillouin Zone (BZ). We label the two valleys as $+$ and $-$. For each valley, the effective low energy model is a simple two band model, consisting of the $A$ sublattice of the top graphene layer and the $B$ sublattice of the bottom graphene layer. Other degrees of freedom are not active at low energy, and can be ignored.  For the valley $+$, the effective model in the basis $(c^t_A,c^b_B)$  is:
\begin{align}
 & h_{+}(\mathbf k)=\notag\\
 &\left(
  \begin{array}{cc}
  \frac{\Delta_V}{2} & \frac{t^3}{\gamma_1^2}(k_x-ik_y)^3+2\frac{t\gamma_3}{\gamma_1}|\mathbf k|^2)\\
  \frac{t^3}{\gamma_1^2}(k_x+ik_y)^3+2\frac{t\gamma_3}{\gamma_1}|\mathbf k|^2) &-\frac{\Delta_V}{2}
  \end{array}
  \right)
  \label{eq:valley_h0}
\end{align}
 We use $t=-3000\frac{\sqrt{3}}{2}$ meV, $\gamma_1=380$ meV and $\gamma_3=293\frac{\sqrt{3}}{2}$ meV. However we do not expect these parameters to be quantitatively precise. In the above equation momentum $\mathbf k$ is in units of $\frac{1}{a}$.  $\Delta_V$ is the energy difference between the top and the bottom graphene layers, which is controlled by an applied voltage. The model for the valley $-$ is the time reversal transformation of the above model.

Then moir\'e lattice gives a super-lattice potentials: 
\begin{equation}
   H_M=\sum_{a;\mathbf{k},\mathbf{G_j}}c_{a;t}^\dagger(\mathbf{k+G_j})V(\mathbf G_j)c_{a;t}(\mathbf{k})+h.c.
   \label{eq:HM}
 \end{equation} 
 where $G_j$ is the moir\'e super-lattice reciprocal vector and $a=+,-$ is the valley index. We choose $\mathbf{G_1}=(0,\frac{4\pi}{\sqrt{3}a_M})$ and $\mathbf{G_2}=(-\frac{2\pi}{\sqrt{3}a_M},\frac{2\pi}{a_M})$ for the moir\'e Brillouin zone (MBZ).  Because only the h-BN on top of the graphene is aligned and effective, we expect the moir\'e superlattice potential only acts on the $c^t_A$ component. We use $V(\mathbf{G_1})=V_0e^{i\theta_0}$ with $V_0=-14.88$ meV
		and $\theta_0=-50.19^\circ$. $V(G_j)$ for other $j$ can be generated by $C_6$ rotation: $V(C_6 G)=V(G)^*$.

\begin{figure}
\centering
\includegraphics[width=0.45\textwidth]{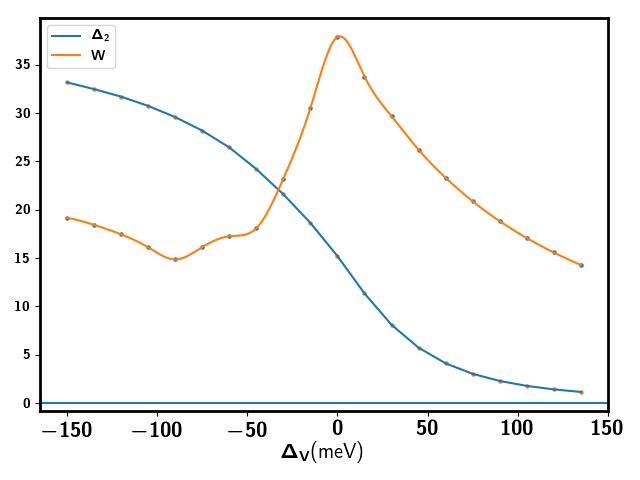}
\caption{The dependence of the band width $W$ and the band gap  $\Delta_2$ on the applied vertical voltage difference $D$. $\Delta_2$ is the minimal gap between the valence band and the band below. $W$, $\Delta_2$ are all in units of meV. The band gap $\Delta_1$ between the conduction band and the valence band is almost equal to $|\Delta_V|$ and becomes larger than the bandwidth after $|\Delta_V|>30$ meV.}
\label{fig:band_gap_weidth}
\end{figure}

The bandwidth can be tuned by $\Delta_V$, as shown in Fig.~\ref{fig:band_gap_weidth}.

 \subsection{Symmetry}
 We first discuss the symmetries of the continuum model  of Eq.~\ref{eq:valley_h0} and Eq.~\ref{eq:HM}.

 First, there is  time reversal symmetry which relates the two valleys: complex conjugation combined with $c_{+,\alpha}(\mathbf k)\rightarrow c_{-,\alpha}(\mathbf k)$ where $\alpha=t,b$ is the spinor index.  Both Eq.~\ref{eq:valley_h0} and Eq.~\ref{eq:HM} are also apparently invariant under $C_3$ rotation symmetry: $c_{a;\alpha}(\mathbf k)\rightarrow c_{a;\alpha}(C_3 \mathbf k)$ where $a=+,-$ is the valley index and $\alpha=t,b$ is spinor index in Eq.~\ref{eq:valley_h0}. There is no inversion symmetry (and therefore $C_6$ rotation symmetry) in Eq.~\ref{eq:valley_h0} and Eq.~\ref{eq:HM}.

 Within the continuum model there is also a mirror reflection symmetry along the $\mathbf{G_6}=(2\pi,\frac{2\pi}{\sqrt{3}})$: $\theta(M\mathbf k)=\frac{\pi}{3}-\theta(\mathbf k)$ where $\theta(\mathbf k)$ is the angle of $\mathbf k$ in the polar coordinate. The Hamiltonian in Eq.~\ref{eq:valley_h0} and Eq.~\ref{eq:HM} is invariant under the Mirror symmetry $c_{+,\alpha}(\mathbf k)\rightarrow c_{-;\alpha}(M\mathbf k)$.  However, microscopically this  mirror reflection should be broken by the h-BN layer. We view it as a a good approximation in the continuum model.

 \subsection{Band Structures in a small out-of-plane magnetic field}
The moir\'e superlattice folds the orginal band of TLG to a moir\'e Brillouin Zone (MBZ) which is a hexagon. We take both valleys of the original band to be the $\Gamma$ point of the MBZ.

\begin{figure}[ht]
\centering
  \begin{subfigure}[b]{0.3\textwidth}
    \includegraphics[width=\textwidth]{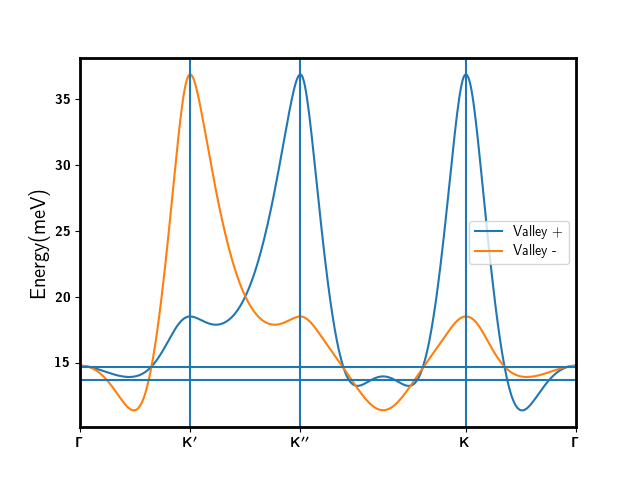}
    \caption{$\Delta_V=-25$ meV, $B=0$ T, $C=0$}
  \end{subfigure}
  \begin{subfigure}[b]{0.33\textwidth}
    \includegraphics[width=\textwidth]{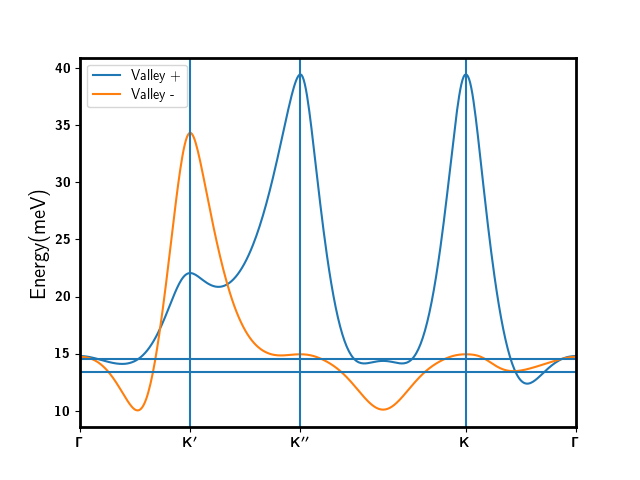}
    \caption{$\Delta_V=-25$ meV, $B=1$ T, $C=0$}
  \end{subfigure}
  \begin{subfigure}[b]{0.33\textwidth}
    \includegraphics[width=\textwidth]{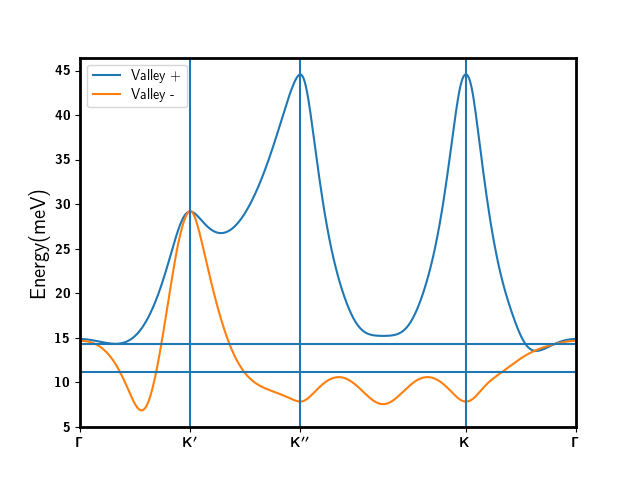}
    \caption{$\Delta_V=-25$ meV, $B=3$T, $C=0$}
    \label{fig:B3D0}
  \end{subfigure}

  \begin{subfigure}[b]{0.33\textwidth}
    \includegraphics[width=\textwidth]{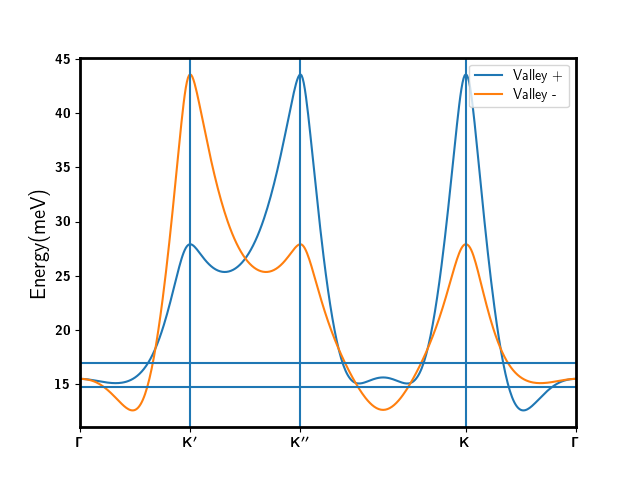}
    \caption{$\Delta_V=25$ meV, $B=0$ T, $C=\pm 3$}
  \end{subfigure}
  \begin{subfigure}[b]{0.3\textwidth}
    \includegraphics[width=\textwidth]{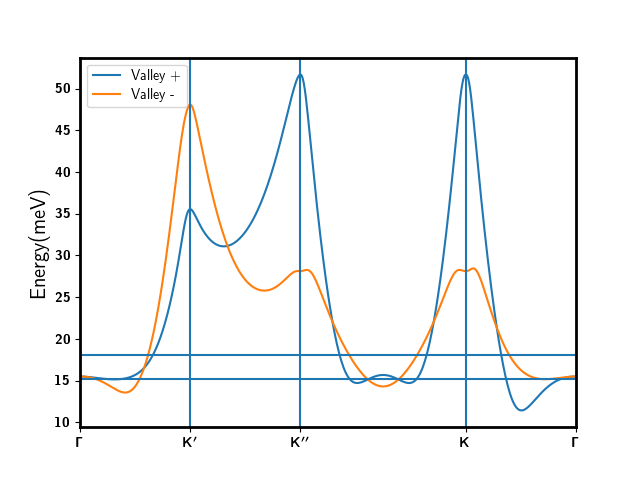}
    \caption{$\Delta_V=25$ meV, $B=1$ T, $C=\pm 3$}
  \end{subfigure}
  \begin{subfigure}[b]{0.3\textwidth}
    \includegraphics[width=\textwidth]{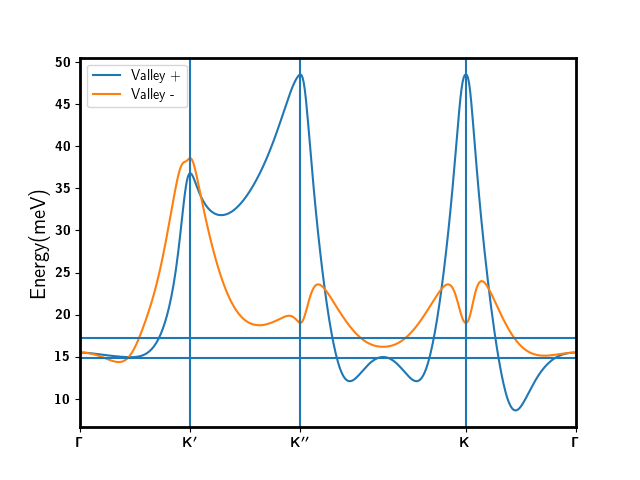}
    \caption{$\Delta_V=25$ meV, $B=3$T, $C=\pm 3$}
  \end{subfigure}

  \caption{Band structures of the valence bands in the hole picture for $\Delta_V=-25$ meV and $\Delta_V=25$ meV in ab out-of-plane magnetic field $B$. $K'=(0,\frac{4\pi}{3a_M})$. $K''=(\frac{2\pi}{\sqrt{3}a_M},\frac{2\pi}{3a_M})$ and $K=(0,-\frac{4\pi}{3a_M})$ are equivalent in the MBZ. Two horizontal lines are the chemical potential for $\nu_T=1$ and $\nu_T=2$.  For $\Delta_V<0$, out-of-plane magnetic field split the energies of two valleys. It also increases the band width of one valley while reducing the band width of the other valley. For $\Delta_V>0$, out-of-plane magnetic field increase the bandwidth of one valley while decrease the bandwidth of the other valley. }
  \label{fig:Band_structures}
\end{figure}

We show band structures of the valence bands for TLG/h-BN system in a small out-of-plane magnetic field in Fig.~\ref{fig:Band_structures} incorporating the effects of the valley Zeeman coupling.

\subsection{Fermi Surfaces at $\nu_T=1,2$ for $\Delta_V<0$}
\begin{figure}[ht]
\centering
  \begin{subfigure}[b]{0.45\textwidth}
    \includegraphics[width=\textwidth]{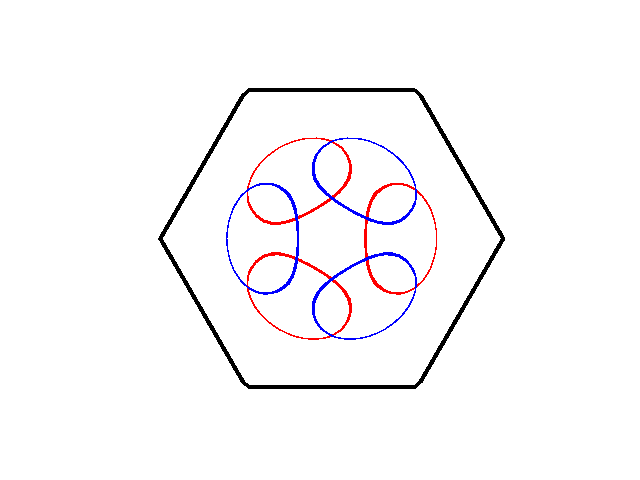}
    \caption{$\Delta_V=-10$ meV, $\nu_T=1$}
  \end{subfigure}
   \begin{subfigure}[b]{0.45\textwidth}
    \includegraphics[width=\textwidth]{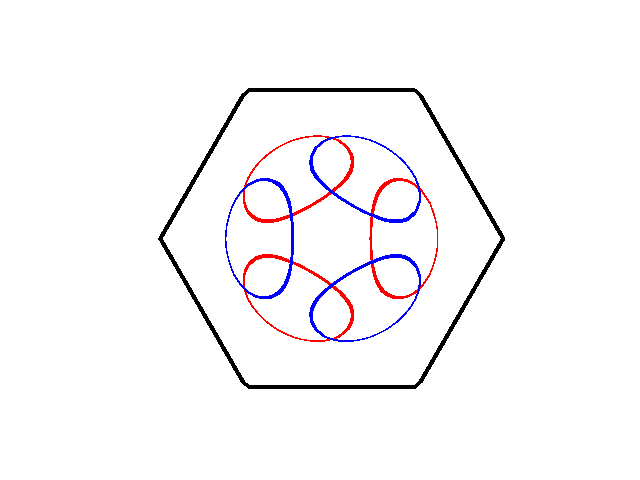}
    \caption{$\Delta_V=-20$ meV, $\nu_T=1$}
  \end{subfigure}

    \begin{subfigure}[b]{0.45\textwidth}
    \includegraphics[width=\textwidth]{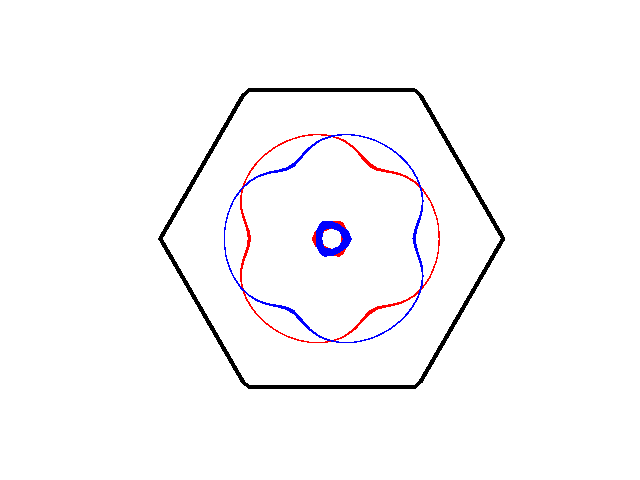}
    \caption{$\Delta_V=-10$ meV, $\nu_T=1.6$}
  \end{subfigure}
  \begin{subfigure}[b]{0.45\textwidth}
    \includegraphics[width=\textwidth]{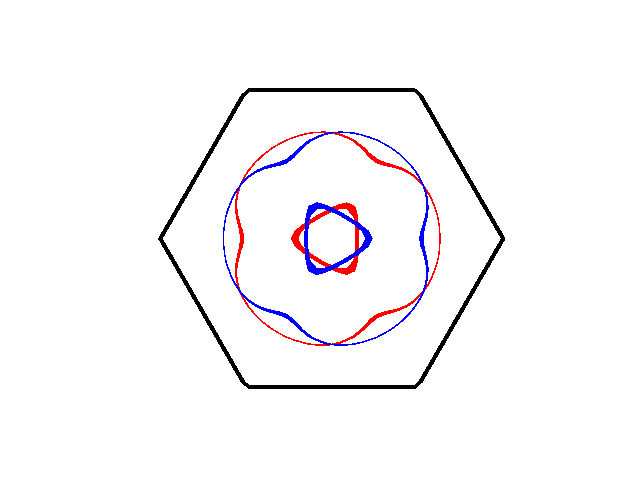}
    \caption{$\Delta_V=-20$ meV, $\nu_T=1.6$}
  \end{subfigure}

   \begin{subfigure}[b]{0.45\textwidth}
    \includegraphics[width=\textwidth]{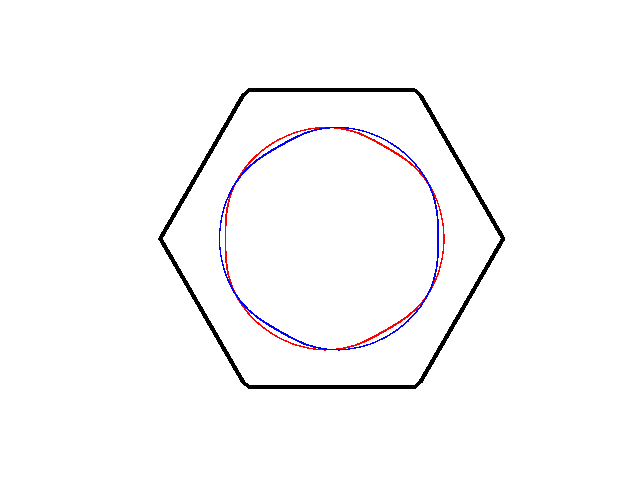}
    \caption{$\Delta_V=-10$ meV, $\nu_T=2$}
  \end{subfigure}
  \begin{subfigure}[b]{0.45\textwidth}
    \includegraphics[width=\textwidth]{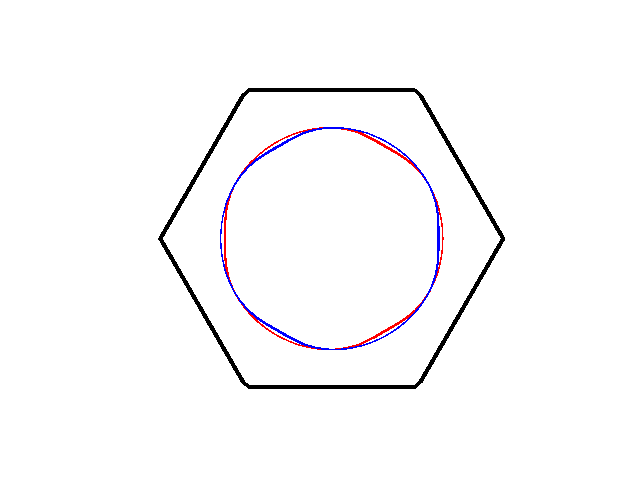}
    \caption{$\Delta_V=-20$ meV, $\nu_T=2$}
  \end{subfigure}

  \caption{ Fermi Surfaces at $\nu_T=1$ and $\nu_T=2$. Red and blue lines denotes the Fermi surface contours for the two different valleys. For $\nu_T=1$, there are three separate Fermi surfaces related by the $C_3$ symmetry. When increasing $\nu_T$, there is a Lifshitz transition to an annulus-shape Fermi sea. At $\nu_T=2$, the Fermi surface is simply a circle for each valley.}
  \label{fig:Fermi_Surface}
\end{figure}

To aid the discussion of the metal-insulator transition for $\nu_T=1,2$ in the $\Delta_V<0$ side, we provide the plots of the Fermi surfaces at several different values of $\Delta_V$ in Fig.~\ref{fig:Fermi_Surface}. In our model, the Fermi surfaces do not have an obvious nesting instability in the particle-hole channel. For $\nu_T=1$, the filled Fermi sea has the topology close to  $\Delta_V \approx -20$ meV.

\section{Hamiltonian In Momentum Space}

In momentum space, we focus on the four valence bands labeled by spin $\sigma=\uparrow,\downarrow$ and valley $a=+,-$.  The density operator projected to the valence bands is
\begin{align}
  \rho(\mathbf x)&=\sum_{a\sigma,\mathbf k, \mathbf q}\lambda_a(\mathbf k,\mathbf q)c^\dagger_{a\sigma}(\mathbf{k+q})c_{a\sigma}(\mathbf k) e^{-i\mathbf{q}\cdot \mathbf{x}}\notag\\
  &+\sum_{\sigma,\mathbf k,\mathbf q}\left(\lambda_{+-}(\mathbf k, \mathbf q)c^\dagger_{+\sigma}(\mathbf{k+q})c_{-\sigma}(\mathbf k)e^{-i (\mathbf{2K+q})\cdot \mathbf{x}}+\lambda_{-+}(\mathbf k, \mathbf q)c^\dagger_{-\sigma}(\mathbf{k+q})c_{+\sigma}(\mathbf k)e^{-i (\mathbf{-2K+q})\cdot \mathbf{x}}\right)
\end{align}
where $\mathbf{K}=(\frac{4\pi}{3 a},0)$, $a=0.236$ nm is the lattice constant of the graphene layer. Form factors $\lambda_{a}(\mathbf k, \mathbf q)$ and $\lambda_{+-}(\mathbf k, \mathbf q)$ can be calculated in the continuum model approach following Ref.~\onlinecite{zhang2018moir}.

The full Hamiltonian is

\begin{align}
  H&=\sum_{\mathbf k;a,\sigma}\xi_a(\mathbf k) c^\dagger_{a\sigma}(\mathbf k)c_{a\sigma}(\mathbf k)\notag\\
  &+\frac{1}{2}\int \frac{d^2\mathbf q}{(2\pi)^2}\sum_{\mathbf{k_1},\mathbf{k_2}; a_1,\sigma_1,a_2,\sigma_2}c^\dagger_{a_1,\sigma_1}(\mathbf{k_1+q})c^\dagger_{a_2,\sigma_2}(\mathbf{k_2-q})c_{a_2,\sigma_2}(\mathbf k_2)c_{a_1,\sigma_1}(\mathbf k_1) V(\mathbf q) \lambda_{a_1}(\mathbf{k_1},\mathbf{q})\lambda_{a_2}(\mathbf{k_2},-\mathbf{q})\notag\\
  &+\frac{1}{2}\int \frac{d^2\mathbf q}{(2\pi)^2}\sum_{\mathbf{k_1},\mathbf{k_2}; \sigma_1\sigma_2}\left(c^\dagger_{+,\sigma_1}(\mathbf{k_1+q})c^\dagger_{-,\sigma_2}(\mathbf{k_2-q})c_{+,\sigma_2}(\mathbf k_2)c_{-,\sigma_1}(\mathbf k_1) V(2\mathbf{K}+\mathbf q) \lambda_{+-}(\mathbf{k_1},\mathbf{q})\lambda_{-+}(\mathbf{k_2},-\mathbf{q})+h.c.\right)
  \label{eq:momentum_hamiltonian}
\end{align}

where we use screened Coulomb interaction $V(\mathbf q)=\frac{e^2}{2 \xi_0 \kappa} \frac{1}{q}(1-e^{- q r_0})$. $\kappa$ is the renormalized factor for dielectric constant. In this paper we use $\kappa=8$. $r_0$ is the screening length for which we use $r_0=5 a_M\approx 75$ nm. 

 The first two terms of Eq.~\ref{eq:momentum_hamiltonian} have $SU(2)_+\times SU(2)_-\times U(1)_v$ symmetry, which means that $SU(2)$ spin of each valley is separately conserved. The third term breaks it further down to $U(1)_c\times U(1)_v\times SU(2)_s$. We expect this term is suppressed by a factor $\frac{a}{a_M}\approx 0.02$ and therefore we only view it as a perturbation.

\section{Lattice Model for $\Delta_V<0$ Side \label{appendix:lattice_models}}

For $\Delta_V<0$, the bands of both valleys are trivial($C=0$). Therefore there is exponentially localized Wannier orbital for each valley created by

\begin{equation}
  \psi^\dagger_a(\mathbf{x_0})=\frac{1}{\sqrt{N}}\sum_{\mathbf k}e^{-i \mathbf{k}\cdot \mathbf{x_0}}e^{i\theta_a(\mathbf k)}c^\dagger_a(\mathbf k)
 \end{equation} 

$\theta_a(\mathbf k)$ can be obtained by the standard projection methodd\cite{marzari2012maximally}: $A=\braket{\mu_a(\mathbf k)|g_a(\mathbf k)}$ and $e^{i\theta_a(\mathbf k)}=A/|A|$. $\mu_a(\mathbf k)$ is the Bloch wave-function for valley $a$. $g_a(\mathbf k)$ is an initial ansatz localized in real space. We choose $g_a(\mathbf k)$ to be the Fourier transformation of 
\begin{equation}
 \ket{ g_a(\mathbf x)}=e^{-\frac{(\mathbf x-\mathbf{x_0})^2}{2 \alpha^2}} \ket{\phi_a}
\end{equation}
where $\alpha=\frac{a_M}{16}$ and $\ket{\phi_a}$ is a constant vector corresponding to sublattices (which can be viewed as pseudospin degree of freedom and for simplicity we assume an ansatz for which the pseudospin is independent of $\mathbf{x}$).  The value of $\ket{\phi_a}$ is chosen to optimize the overlap $|\braket{\mu_a(\mathbf k)|g_a(\mathbf k)}|$.

After getting $\theta_a(\mathbf k)$, we can easily transform the Hamiltonian in Eq.~\ref{eq:momentum_hamiltonian} in terms of real space operator $\psi_a(\mathbf{x})$ with $\mathbf{x}=\mathbf{x_0}+m \mathbf{a_1}+n \mathbf {a_2}$ forming a two dimensional triangular lattice. $\mathbf{a_1}=a_M(1,0)$ and $\mathbf{a_2}=a_M(\frac{1}{2},\frac{\sqrt{3}}{2})$.

For the kinetic term,  we have
\begin{equation}
  H_K=-\sum_{a;m,n,\mathbf{x}}t_a(m,n)\psi_a^\dagger(\mathbf x +m \mathbf{a_1}+n\mathbf{a_2})\psi_a(\mathbf x)+h.c.
 \end{equation}
 with 
 \begin{equation}
  t_a(m,n)=-\frac{1}{N}\sum_{\mathbf k}\xi_a(\mathbf k)e^{- i \mathbf{k}\cdot(m \mathbf{a_1}+n \mathbf{a_2})}
 \end{equation}

 Similarly we can generate all of four fermion interactions. The second line in Eq.~\ref{eq:momentum_hamiltonian} gives
 \begin{equation}
  H_V=\frac{1}{2}\sum_{\mathbf{x},\mathbf{R_1},\mathbf{R_2},\mathbf{R_3}}V_{ab}(\mathbf{R_1},\mathbf{R_2},\mathbf{R_3})\psi^\dagger_{a\sigma_1}(\mathbf x)\psi^\dagger_{b\sigma_2}(\mathbf{x+R_1})\psi_{b\sigma_2}(\mathbf{x+R_2})\psi_{a\sigma_1}(\mathbf{x+R_3})
\end{equation}
with
\begin{align}
  &\ \ \ V_{ab}(\mathbf{R_1},\mathbf{R_2},\mathbf{R_3})\notag\\
  &=\frac{1}{N^3}\sum_{\mathbf{k_1},\mathbf{k_2},\mathbf{q}}\sum_{a,b}V(\mathbf q)e^{-i \theta_a(\mathbf{k_1+q})}e^{-i \theta_b(\mathbf{k_2-q})}e^{i \theta_b(\mathbf{k_2})}e^{i \theta_a(\mathbf{k_1})}\lambda_a(\mathbf{k_1},\mathbf{q})\lambda_b(\mathbf{k_2},-\mathbf{q}) e^{i(\mathbf{k_2-q})\cdot \mathbf{R_1}}e^{-i \mathbf{k_2}\cdot \mathbf{R_2}}e^{-i \mathbf{k_1}\cdot \mathbf{R_3}}
\end{align}

The dominant term is onsite and nearest neighbor Hubbard $U$. The next order is Hund's coupling, as shown in Eq.~\ref{eq:trivial_interaction}. There are also pair hopping and correlated hopping terms:

\begin{equation}
  \sum_{ab}\left(g_{dh}U\psi^\dagger_{a\sigma_1}(\mathbf x)\psi^\dagger_{b\sigma_2}(\mathbf x+\mathbf{a_1})\psi_{b\sigma_2}(\mathbf x)\psi_{a\sigma_1}(\mathbf x)+
  g_{hh}U\psi^\dagger_{a\sigma_1}(\mathbf x)\psi^\dagger_{b\sigma_2}(\mathbf x+\mathbf{a_1})\psi_{b\sigma_2}(\mathbf x)\psi_{a\sigma_1}(\mathbf x+\mathbf{a_1})+h.c.\right)
\end{equation}

These terms $g_{dh}U\sim g_{hh}U\sim0.02 U\approx 0.5$ meV, which is at the same order of Hund's coupling $g_h U$ term in Eq.~\ref{eq:trivial_interaction}. However, they involve onsite double occupancy, which should be suppressed by the much larger Hubbard $U$. Therefore as a simplifying approximation we only keep Hund's coupling term and ignore these correlated hopping and pair hopping terms.

Last, we also need to include the third line of Eq.~\ref{eq:momentum_hamiltonian}. It turns out that in real space this terms leads to an onsite inter-valley Hund's coupling, i.e. $J_H$ term in Eq.~\ref{eq:trivial_interaction}. 

\section{$\Delta_V>0$: $C=\pm 3$ Chern Bands\label{appendix:Chern-lattice-models}}
For $\Delta_V>0$, localized Wannier orbitals for each valley are impossible because of the non-zero Chern number. However, we can have a triangular lattice model with two orbitals per site at the cost that the valley $I_z$ operator can not be on-site. This kind of model was first discussed for the topologically non-trivial bands of the twisted bilayer graphene system\cite{po2018origin}.

We choose two initial localized ansatz $\ket{g_1}$ and $\ket{g_2}$ centered at a triangular lattice. They are related by the time reversal transformation. We label valley $+,-$ as $1,2$. Then we calculate the following $2\times 2$ matrix: $A_{mn}(\mathbf k)=\braket{\mu_m(\mathbf k)|g_n(\mathbf k)}$ at each momentum point $\mathbf k$. This give the following two projected states:
\begin{align}
\varphi_n^\dagger(\mathbf x_0)=\frac{1}{\sqrt{N}}\sum_{\mathbf k;a} A_{an}(\mathbf k) e^{-i \mathbf{k}\cdot \mathbf{x_0}}c_a^\dagger(\mathbf k)
\end{align}
where $m,n=1,2$ and $c_a(\mathbf k)$ is the annilation operator of the valley $a$.

$\varphi_{1}$ and $\varphi_{2}$ create states which are not orthogonal and normalized. We define the following unitary matrix:
\begin{equation}
  U(\mathbf k)=A (A^\dagger A)^{-\frac{1}{2}}
\end{equation}

If we do Singular Value Decomposition: $A=ZDW^\dagger$, then
\begin{equation}
  U(\mathbf k)=Z W^\dagger
\end{equation}

Then we get Wannier orbitals:
\begin{align}
\psi_n^\dagger(\mathbf x_0)=\frac{1}{\sqrt{N}}\sum_{\mathbf k;a}  e^{-i \mathbf{k}\cdot \mathbf{x_0}}c_a^\dagger(\mathbf k)U_{an}(\mathbf k)
\end{align}
It is easy to prove that $\ket{\psi_1}=\psi_1^\dagger(\mathbf x_0) \ket{0}$ and $\ket{\psi_2}=\psi_2^\dagger(\mathbf x_0)\ket{0}$ are normalized and orthogonal.  

We can build our lattice models in terms of operators $\psi_1(\mathbf{x_0}+m \mathbf{a_1}+n\mathbf{a_2})$ and $\psi_2(\mathbf{x_0}+m \mathbf{a_1}+n\mathbf{a_2})$ where $\mathbf{a_1}$ and $\mathbf{a_2}$ are the unit vectors of the corresponding triangular lattice.

Using
\begin{equation}
  c_a(\mathbf k)=\frac{1}{\sqrt{N}} \sum_{\mathbf{x_0};n} e^{-i \mathbf{k}\cdot \mathbf{x_0}} U_{an}(\mathbf k)\psi_n(\mathbf x_0)
  \label{eq:c_intermsof_wannier}
\end{equation}
we can express Eq.~\ref{eq:momentum_hamiltonian} in terms of these $\psi_m$ Wannier operators.

First for kinetic term, we have
\begin{align}
H_K&=-\sum_{i,j}\sum_{m,n}t_{mn}(\mathbf R)\psi_{i;m}^\dagger \psi_{j;n}
\end{align}
where $\mathbf R=\mathbf{x_j}-\mathbf{x_i}$. 
\begin{equation}
  t_{mn}(\mathbf R)=\frac{1}{N}\sum_{\mathbf k;a} U^\dagger_{ma}(\mathbf k)\xi_a(\mathbf k) U_{an}(\mathbf k)e^{-i\mathbf{k}\cdot \mathbf{R}}
\end{equation}
where $m,n=A,B$ are orbital indexes for each site. $a=+,-$ is valley index. $i,j$ are labels of triangular lattice sites.

We keep intra-orbital hopping $t(x,y)=t_{AA}(x\mathbf{a_1}+y\mathbf{a_2})$ and inter-orbital hopping $t'(x,y)=t_{AB}(x \mathbf{a_1}+y\mathbf{a_2})$. Other components can be generated by the time reversal transformation: $\psi_{m;i}\rightarrow \epsilon_{mn}\psi_{n;i}$, where $\epsilon_{AB}=-\epsilon_{BA}=1$ while $\epsilon_{AA}=\epsilon_{BB}=0$.

There is always the following symmetry $t(\mathbf x)=t(C_3 \mathbf{x})=t^*(-\mathbf x)=t^*(C_6 \mathbf{x})$ and $t'(\mathbf x)=t'(C_6 \mathbf{x})$. The mirror reflection symmetry can not be kept explicitly in the current approach.

$I_z=\sum_{\mathbf k} c^\dagger_{+}(\mathbf k)c_{+}(\mathbf k)-c^\dagger_{-}(\mathbf k)c_{-}(\mathbf k)$ can not be implemented as on site operator in the Wannier orbital $\psi_{i;m}$ basis.  Instead, we have
\begin{equation}
  I_z=\sum_{i,j}\sum_{m,n}t^v_{mn}(\mathbf R)\psi_{i;m}^\dagger \psi_{j;n}
\end{equation}

Again we have intra-orbital hopping $t_V(x,y)=t_{AA}(x\mathbf{a_1}+y\mathbf{a_2})$ and inter-orbital hopping $t'_V(x,y)=t_{AB}(x \mathbf{a_1}+y\mathbf{a_2})$.  The symmetry requirement is: $t_V(\mathbf x)=t_V(C_6 \mathbf{x})$ and $t'_V(\mathbf x)=-t^{'}_V(C_6 \mathbf{x})$.  $t_{BB}(\mathbf x)=- t_{AA}(\mathbf x)$ and  $t_{AB}(\mathbf x)=-t^*_{BA}(\mathbf x)$ follow from $T I_z T^{-1}=-I_z$ under time reversal.

Similarly to the $\Delta_V<0$ case, four fermion interaction can be expressed in terms of Wannier orbital operator $\psi_{i;m}$ in real space:

\begin{equation}
  H_V=\frac{1}{2}\sum_{\sigma_1;\sigma_2}\sum_{\mathbf{x},\mathbf{R_1},\mathbf{R_2},\mathbf{R_3}}V_{m_1n_1n_2m_2}(\mathbf{R_1},\mathbf{R_2},\mathbf{R_3})\psi^\dagger_{m_1\sigma_1}(\mathbf x)\psi^\dagger_{n_1\sigma_2}(\mathbf{x+R_1})\psi_{n_2\sigma_2}(\mathbf{x+R_2})\psi_{m_2\sigma_1}(\mathbf{x+R_3})
\end{equation}
where,
\begin{align}
  &\ \ \ V_{m_1n_1n_2m_2}(\mathbf{R_1},\mathbf{R_2},\mathbf{R_3})\notag\\
  &=\frac{1}{N^3}\sum_{\mathbf{k_1},\mathbf{k_2},\mathbf{q}}\sum_{a,b}V(\mathbf q)U^*_{am_1}(\mathbf{k_1+q})U^{*}_{bn_1}(\mathbf{k_2-q})U_{bn_2}(\mathbf k_2)U_{am_2}(\mathbf k_1)\lambda_a(\mathbf{k_1},\mathbf{q})\lambda_b(\mathbf{k_2},-\mathbf{q}) e^{i(\mathbf{k_2-q})\cdot \mathbf{R_1}}e^{-i \mathbf{k_2}\cdot \mathbf{R_2}}e^{-i \mathbf{k_1}\cdot \mathbf{R_3}}
  \label{eq:Chern_interaction_convention}
\end{align}

\subsection{Result}

We provide a two-orbital model for the $C=\pm 3$ bands following the procedure described above. For simplicity we ignore the trigonal warping term $\gamma_3$ of Eq.~\ref{eq:valley_h0} for the calculation of $\Delta_V>0$.

Tight binding parameters for $H_K$ and $I_z$ are listed in Table.~\ref{table:tight_binding_parameters} and in Table.~\ref{table:Iz_tight_binding_parameters} for $\Delta_V=50$ meV. These tight binding parameters for a two orbital model can reproduce the two valence bands for each spin with Chern number $C=\pm 3$.

\begin{table}[ht]
\centering
\begin{tabular}{c|c|c|c|c|c|c}

\hline
$\mathbf{R}$&$(0,0)$&$(1,0)$&$(1,1)$&$(2,1)$&$(1,2)$&$(2,0)$\\
\hline
$t$&$0$&$0.984e^{i0.006\pi}$& $-0.836$& $0.293e^{-i0.002\pi}$&$0.293e^{i0.002\pi}$&$-0.236e^{i0.021\pi}$\\
\hline
$t'$&$-0.155e^{-i0.380\pi}$&$0.064e^{-i0.380\pi}$ & $-0.082e^{-i0.379\pi}$& $0.026e^{-i0.256\pi}$&$-0.026e^{i0.496\pi}$&$0.027e^{-i0.378\pi}$\\
\hline
\end{tabular}
\caption{Tight binding parameters  of  the kinetic Hamiltonian $H_K$ for $\Delta_V=50$ meV. $t$ and $t'$ are intra-orbital and inter-orbital hopping parameters in units of meV.}
\label{table:tight_binding_parameters}
\end{table}

\begin{table}[ht]
\centering
\begin{tabular}{c|c|c|c|c|c|c}

\hline
$\mathbf{R}$&$(0,0)$&$(1,0)$&$(1,1)$&$(2,1)$&$(1,2)$&$(2,0)$\\
\hline
$t_V$ &$0.347$& $0.189$& $-0.133$&$0.020$&$0.019$&$-0.090$\\
\hline
$t'_V$&$0$&$0.252e^{i0.121\pi}$ & $-0.172e^{-i0.380\pi}$& $-0.044e^{i0.493\pi}$&$0.044e^{-i0.255\pi}$&$0.038e^{i0.125\pi}$\\
\hline
\end{tabular}
\caption{Tight binding parameters  of the valley operator $I_z$ for $\Delta_V=50$ meV. $t$ and $t'$ are intra-orbital and inter-orbital hopping parameters in units of meV.}
\label{table:Iz_tight_binding_parameters}
\end{table}

\begin{table}[ht]
\centering
\begin{tabular}{c|c|c|c|c|c|c|c|c}

\hline
$m_1$&$n_1$&$n_2$&$m_2$&$\mathbf{R_1}$&$\mathbf{R_2}$&$\mathbf{R_3}$&$V_{m_1n_1n_2m_2}(\mathbf{R_1,\mathbf{R_2},\mathbf{R_3}})$ &Comments\\
\hline
A &A &A &A &$(0,0)$& $(0,0)$ &$(0,0)$& $9.39$& on-site U\\
A &A &A &A &$(1,0)$& $(1,0)$ &$(0,0)$& $4.09$&nearest neighbor U\\
A &A &A &A &$(1,0)$& $(0,0)$ &$(0,0)$& $1.08e^{-i0.008 \pi}$& correlated hopping \\
A &A &A &A &$(1,0)$& $(0,0)$ &$(1,0)$& $0.65$ & inter-site Hund's\\
A &A &A &A &$(1,0)$& $(1,0)$ &$(0,0)$& $0.50$ & pair Hopping\\
\hline
A &B &B &A &$(0,0)$& $(0,0)$ &$(0,0)$& $9.39$& on-site U\\
A &B &B &A &$(1,0)$& $(1,0)$ &$(0,0)$& $4.09$&nearest neighbor U\\
A &B &B &A &$(1,0)$& $(0,0)$ &$(0,0)$& $1.08e^{i0.008 \pi}$& correlated hopping \\
A &B &B &A &$(1,0)$& $(0,0)$ &$(1,0)$& $0.50$ & inter-site Hund's\\
A &B &B &A &$(1,0)$& $(1,0)$ &$(0,0)$& $0.65$ & pair Hopping\\
\hline
A &B &A &B &$(0,0)$& $(0,0)$ &$(0,0)$& $0.15$& \\
A &B &A &B &$(1,0)$& $(1,0)$ &$(0,0)$& $0.017e^{i0.77\pi}$&\\
A &B &A &B &$(1,0)$& $(0,0)$ &$(0,0)$& $0.07e^{i0.42 \pi}$& \\
A &B &A &B &$(1,0)$& $(0,0)$ &$(1,0)$& $0.73$ & \\
A &B &A &B &$(1,0)$& $(1,0)$ &$(0,0)$& $0.73$ & \\
\hline
A &A &A &B &$(0,0)$& $(0,0)$ &$(0,0)$& $0.55 e^{i0.58 \pi}$& \\
A &A &A &B &$(1,0)$& $(1,0)$ &$(0,0)$& $0.16e^{i0.68\pi}$&\\
A &A &A &B &$(1,0)$& $(0,0)$ &$(0,0)$& $0.07e^{i0.70 \pi}$& \\
A &A &A &B &$(1,0)$& $(0,0)$ &$(1,0)$& $0.13e^{0.018\pi}$ & \\
A &A &A &B &$(1,0)$& $(1,0)$ &$(0,0)$& $0.147e^{-i0.78\pi}$ & \\
\hline
A &A &B &A &$(0,0)$& $(0,0)$ &$(0,0)$& $0.55 e^{i0.58 \pi}$& \\
A &A &B &A &$(1,0)$& $(1,0)$ &$(0,0)$& $0.16e^{i0.54\pi}$&\\
A &A &B &A &$(1,0)$& $(0,0)$ &$(0,0)$& $0.79e^{i0.078 \pi}$& \\
A &A &B &A &$(1,0)$& $(0,0)$ &$(1,0)$& $0.147e^{-i0.78 \pi}$ & \\
A &A &B &A &$(1,0)$& $(1,0)$ &$(0,0)$& $0.147e^{-i0.78\pi}$ & \\
\hline
\end{tabular}
\caption{Interaction parameters (in units of meV) for $\Delta_V=50$ meV. The list is not complete. Other terms can be generated from Hermitian conjugation and time reversal transformation.}
\label{table:chern_interaction_parameters}
\end{table}

One can also write four fermion interactions in terms of these $\psi_m$ operators.  We list the dominant interaction terms following the convention of Eq.~\ref{eq:Chern_interaction_convention} in Table.~\ref{table:chern_interaction_parameters}.  The dominant term is still the on-site $U$ and the next-nearest-neighbor repulsion $U_1$. For the $\Delta_V>0$ side $U=10$ meV and is only one half of the value at the $\Delta_V<0$ side. Meanwhile the inter-site Hund's coupling and correlated hopping terms are at the order of $0.1 U$ instead of $0.01 U$ for the $\Delta_V<0$ side. These are signatures of the Wannier obstruction.  In the $\frac{t}{U}<<0$ limit for integer fillings, we still expect an insulating ground state. These inter-site Hund's coupling, correlated hopping and pair-hopping terms are much larger than the super-exchange $\frac{t^2}{U}$ terms and we expect the ground state is decided by these terms. However, the lack of the explicit valley index makes it hard to reliably deal with this lattice model. From Hartree Fock calculations in the momentum space \cite{zhang2018moir} we expect the ground state to be valley polarized for $\nu_T=1$. But we do not know how to understand this conclusion from the above lattice model.

\section{Spin-Valley model for $C=0$ side \label{appendix:spin-valley model}}
For the $C=0$ side, to order $t<<U$, we derive a spin-valley model following the standard approach. There is already a Hund's coupling in the four fermion interaction. Besides, at the order  of $t^2/U$ we get the following super-exchange antiferromagnetic term:
\begin{equation}
  H_S=\frac{t^2}{U}\sum_{\langle ij \rangle} \sum_{a_1\sigma_1, a_2 \sigma_2}\left(e^{i (\varphi^{ij}_{a_1}-\varphi^{ij}_{a_2})}f^\dagger_{i;a_1\sigma_1}f_{i;a_2\sigma_2} f^\dagger_{j;a_2\sigma_2}f_{j;a_1\sigma_1}+h.c.\right)
\end{equation}
where $a_1,a_2$ are valley indexes and $\sigma_1,\sigma_2$ are spin indexes. We use $f$ instead of $c$ to emphasize that they are neutral degrees of freedom which only carry spin and valley quantum numbers. $\varphi_a$ is the phase in the nearest neighbor hopping for valley $a$. From time reversal $\varphi^{ij}_{+}=-\varphi^{ij}_{-}=\varphi$.

 We label operator $\mathbf{\tau_i}\otimes\mathbf{\sigma_i}=f^\dagger_{i;a_1\sigma_1}\mathbf{\tau}_{a_1a_2}\mathbf{\sigma}_{\sigma_1\sigma_2}f_{i;a_2\sigma_2}$ with Einstein summation convention. $\tau_i$ labels $I,\tau_x,\tau_y,\tau_z$ operator acting on the valley Hilbert space at site $i$. Similarly $\sigma_i$ labels $I,\sigma_x,\sigma_y,\sigma_z$.

For $a_1=a_2=+$ part, we use the following equation:
\begin{equation}
  \sum_{\sigma_1\sigma_2}f^\dagger_{i;+\sigma_1}f_{i;+\sigma_2} f^\dagger_{j;+\sigma_2}f_{j;+\sigma_1}=\frac{1}{2}\frac{I+\tau^z_i}{2}\frac{I+\tau^z_j}{2}(I+\mathbf{\sigma_i}\cdot \mathbf{\sigma_j})
\end{equation}
where terms like $\tau \sigma$  refer to tensor products.

Similar for $a_1=a_2=-$ part, we have
\begin{equation}
  \sum_{\sigma_1\sigma_2}f^\dagger_{i;-\sigma_1}f_{i;-\sigma_2} f^\dagger_{j;-\sigma_2}f_{j;-\sigma_1}=\frac{1}{2}\frac{I-\tau^z_i}{2}\frac{I-\tau^z_j}{2}(I+\mathbf{\sigma_i}\cdot \mathbf{\sigma_j})
\end{equation}

Then $a_1=+,a_2=-$ part gives
\begin{equation}
  e^{2i\varphi}\sum_{\sigma_1\sigma_2}f^\dagger_{i;+\sigma_1}f_{i;-\sigma_2} f^\dagger_{j;-\sigma_2}f_{j;+\sigma_1}=\frac{1}{2}e^{2i\varphi}\tau^+_i\tau^{-}_j(1+\mathbf{\sigma_i}\cdot \mathbf{\sigma_j})
\end{equation}
Similarly $a_1=-,a_2=+$ part gives
\begin{equation}
  e^{-2i\varphi}\sum_{\sigma_1\sigma_2}f^\dagger_{i;-\sigma_1}f_{i;+\sigma_2} f^\dagger_{j;-\sigma_2}f_{j;+\sigma_1}=\frac{1}{2}e^{-2i\varphi}\tau^-_i\tau^{+}_j(1+\mathbf{\sigma_i}\cdot \mathbf{\sigma_j})
\end{equation}

Summing the above four terms together, we get the spin-valley coupling from the super-exchange:
\begin{equation}
  \frac{t^2}{2U}\sum_{\langle ij \rangle}\left((I+\mathbf{\tau_i}\cdot \mathbf{\tau_j})(I+\mathbf{\sigma_i}\cdot \mathbf{\sigma_j})-(1-\cos2\varphi_{ij})(\tau^x_i\tau^x_j+\tau^y_i\tau^y_j)(1+\mathbf{\sigma_i}\cdot \mathbf{\sigma_j})+\sin 2\varphi_{ij}(\tau^x_i\tau^y_j-\tau^y_i\tau^x_j)(I+\mathbf{\sigma_i}\cdot \mathbf{\sigma_j})\right)
\end{equation}
where the second and the third term break $SU(4)$ symmetry to $SU(2)_+\times SU(2)_-\times U(1)_v$.
\end{document}